\PassOptionsToPackage{unicode}{hyperref}
\PassOptionsToPackage{hyphens}{url}
\PassOptionsToPackage{dvipsnames,svgnames,x11names}{xcolor}
\documentclass[
  12pt]{article}

\usepackage{amsmath,amssymb}
\usepackage{iftex}
\usepackage{subcaption}
\usepackage{dsfont}
\usepackage{booktabs}
\usepackage{pifont}

\usepackage{longtable}
\newtheorem{theorem}{Theorem}[section]

\newtheorem{proposition}[theorem]{Proposition}
\newtheorem{remark}{Remark}[section]
\numberwithin{equation}{section}
\newtheorem{definition}{Definition}[section]
\newtheorem{assumption}{Assumption}[section]
\ifPDFTeX
  \usepackage[T1]{fontenc}
  \usepackage[utf8]{inputenc}
  \usepackage{textcomp} 
\else 
  \usepackage{unicode-math}
  \defaultfontfeatures{Scale=MatchLowercase}
  \defaultfontfeatures[\rmfamily]{Ligatures=TeX,Scale=1}
\fi
\usepackage{lmodern}
\ifPDFTeX\else  
\fi
\IfFileExists{upquote.sty}{\usepackage{upquote}}{}
\IfFileExists{microtype.sty}{
  \usepackage[]{microtype}
  \UseMicrotypeSet[protrusion]{basicmath} 
}{}
\makeatletter
\@ifundefined{KOMAClassName}{
  \IfFileExists{parskip.sty}{%
    \usepackage{parskip}
  }{
    \setlength{\parindent}{0pt}
    \setlength{\parskip}{6pt plus 2pt minus 1pt}}
}{
  \KOMAoptions{parskip=half}}
\makeatother
\usepackage{xcolor}
\makeatletter
\ifx\paragraph\undefined\else
  \let\oldparagraph\paragraph
  \renewcommand{\paragraph}{
    \@ifstar
      \xxxParagraphStar
      \xxxParagraphNoStar
  }
  \newcommand{\xxxParagraphStar}[1]{\oldparagraph*{#1}\mbox{}}
  \newcommand{\xxxParagraphNoStar}[1]{\oldparagraph{#1}\mbox{}}
\fi
\ifx\subparagraph\undefined\else
  \let\oldsubparagraph\subparagraph
  \renewcommand{\subparagraph}{
    \@ifstar
      \xxxSubParagraphStar
      \xxxSubParagraphNoStar
  }
  \newcommand{\xxxSubParagraphStar}[1]{\oldsubparagraph*{#1}\mbox{}}
  \newcommand{\xxxSubParagraphNoStar}[1]{\oldsubparagraph{#1}\mbox{}}
\fi
\makeatother

\usepackage{longtable,booktabs,array}
\usepackage{calc} 
\usepackage{etoolbox}
\makeatletter
\patchcmd\longtable{\par}{\if@noskipsec\mbox{}\fi\par}{}{}
\makeatother
\IfFileExists{footnotehyper.sty}{\usepackage{footnotehyper}}{\usepackage{footnote}}
\makesavenoteenv{longtable}
\usepackage{graphicx}
\makeatletter
\def\maxwidth{\ifdim\Gin@nat@width>\linewidth\linewidth\else\Gin@nat@width\fi}
\def\maxheight{\ifdim\Gin@nat@height>\textheight\textheight\else\Gin@nat@height\fi}
\makeatother
\setkeys{Gin}{width=\maxwidth,height=\maxheight,keepaspectratio}
\makeatletter
\def\fps@figure{htbp}
\makeatother

\makeatletter
\@ifpackageloaded{caption}{}{\usepackage{caption}}
\AtBeginDocument{%
\ifdefined\contentsname
  \renewcommand*\contentsname{Table of contents}
\else
  \newcommand\contentsname{Table of contents}
\fi
\ifdefined\listfigurename
  \renewcommand*\listfigurename{List of Figures}
\else
  \newcommand\listfigurename{List of Figures}
\fi
\ifdefined\listtablename
  \renewcommand*\listtablename{List of Tables}
\else
  \newcommand\listtablename{List of Tables}
\fi
\ifdefined\figurename
  \renewcommand*\figurename{Figure}
\else
  \newcommand\figurename{Figure}
\fi
\ifdefined\tablename
  \renewcommand*\tablename{Table}
\else
  \newcommand\tablename{Table}
\fi
}
\@ifpackageloaded{float}{}{\usepackage{float}}
\floatstyle{ruled}
\@ifundefined{c@chapter}{\newfloat{codelisting}{h}{lop}}{\newfloat{codelisting}{h}{lop}[chapter]}
\floatname{codelisting}{Listing}

\makeatother
\makeatletter
\@ifpackageloaded{caption}{}{\usepackage{caption}}
\@ifpackageloaded{subcaption}{}{\usepackage{subcaption}}
\makeatother

\ifLuaTeX
  \usepackage{selnolig}  
\fi
\usepackage[]{natbib}
\usepackage{bookmark}

\IfFileExists{xurl.sty}{\usepackage{xurl}}{} 
\hypersetup{
  pdftitle={Invariant quantile regression for heterogeneous environments},
  pdfauthor={Bo Fu; Dandan Jiang},
  pdfkeywords={Causality; Heterogeneous environments; Invariance; Quantile regression; Robust generalization},
  colorlinks=true,
  linkcolor={blue},
  filecolor={Maroon},
  citecolor={Blue},
  urlcolor={Blue},
  pdfcreator={LaTeX via pandoc}}

\newcommand{\anon}{1}

\begin{document}

\def\spacingset#1{\renewcommand{\baselinestretch}%
{#1}\small\normalsize} \spacingset{1}


\if1\anon
{
  \title{\bf Invariant quantile regression for heterogeneous environments}
  \author{Bo Fu\hspace{.2cm}\\
    School of Mathematics and Statistics, Xi'an Jiaotong University\\
    and \\
    Dandan Jiang\thanks{Corresponding author: Dandan Jiang. Email address: jiangdd@xjtu.edu.cn} 
    \\
    School of Mathematics and Statistics, Xi'an Jiaotong University}
  \maketitle
} \fi

\if0\anon
{
  \bigskip
  \bigskip
  \bigskip
  \begin{center}
    {\LARGE\bf Invariant quantile regression for heterogeneous environments}
\end{center}
  \medskip
} \fi

\bigskip
\begin{abstract}
Invariant learning across heterogeneous environments has largely focused on conditional means, which may miss stable causal relationships that arise only at particular parts of the outcome distribution. We develop invariant quantile regression (IQR) for identifying quantile-specific relationships that remain stable across environments while excluding environment-dependent spurious variables. We show that the invariant parameter admits a robust out-of-sample interpretation: it uniquely minimizes the worst-case quantile risk over distribution shifts that preserve the invariant conditional quantile structure. Quantile-based invariance can also yield a richer endogenous set than mean-based invariance, providing additional information for eliminating non-causal variables. To estimate the invariant parameter, we introduce a Kernel-Smoothed Invariant Quantile Regression (KS-IQR) estimator that exploits the underlying invariance structure and heterogeneity among environments. For global minimizers over a bounded parameter space, we establish a non-asymptotic $\ell_2$ error bound and causal discovery (variable selection) properties of our method, and demonstrate its ability to overcome the ``curse of endogeneity'', all in a non-asymptotic framework. We apply our method to real data for causal discovery and obtain biologically meaningful relationships, recovering known signaling pathways and revealing additional quantile-specific effects.
\end{abstract}

\noindent%
\vfill

\newpage
\spacingset{1.8} 

\spacingset{1.8} 

\section{Introduction}\label{sec1}
Data collected from modern scientific studies are often heterogeneous. For example, in large-scale biomedical studies, measurements of the same molecular system may be obtained under different experimental conditions, interventions, laboratories, or batches, leading to substantial distributional variation across data sources \citep{vcuklina2021diagnostics}. Moreover, scientifically relevant effects may not be homogeneous across the outcome distribution: a covariate can have little impact on the average response while exerting pronounced effects in the lower or upper tails \citep{koenker1978regression}. A central statistical problem is therefore to identify, across different parts of the outcome distribution, relationships that remain stable across environments, and to distinguish them from environment-specific associations that may improve in-sample prediction but lead to unstable generalization and scientific conclusions.

The invariance principle provides a natural way to address the environmental heterogeneity by assuming that, although the joint distribution may vary across environments, the causal relationship of interest remains stable. Most existing approaches formulate such invariance through the conditional mean, which seeks a subset of covariates $S$ such that $\mathbb{E}[Y|X_S]$ remains stable across environments \citep{fan2024environment,gu2025causality,gu2025fundamental}. Such methods can exploit environmental heterogeneity to distinguish stable variables from spurious ones. However, conditional-mean invariance is inherently limited because it only captures average behaviors between the covariates and the response. It may therefore fail to detect causal relationships that emerge only in particular regions of the outcome distribution, especially in the tails. In particular, both the magnitude and the support set of covariate effects may vary across quantile levels, so that some covariates can have negligible average effects but exert substantial influence in the tails. For example, \citet{chuang2009causality} point out that causal relationships may manifest only at specific quantile levels and remain undetectable when analysis is restricted to conditional means. In their study of dynamic stock return-volume relations, traditional mean-based causality fails to identify significant causal effects, while quantile-based causality uncovers pronounced causal effects across different quantiles, with particularly strong effects in the tails. Similar motivations also underlie the work of \citet{jeong2012consistent} for the test of Granger causality in quantiles. These findings suggest that causal mechanisms can be inherently heterogeneous across the outcome distribution, and may be obscured by focusing solely on the mean. Another limitation is that existing mean-based methods often rely on sub-Gaussian conditions on the noise or response, which can be restrictive for biomedical and financial data exhibiting heavy-tailed behavior or extreme outliers. Moreover, the endogenous set in these works is defined based on the covariance between the covariates and the noise, which may not fully capture the endogenous structure.

These limitations call for a unified framework that extends the invariance principle beyond the conditional mean. Quantile regression \citep{koenker1978regression} provides a natural framework for studying such distributional heterogeneity by characterizing covariate effects at different parts of the conditional outcome distribution. This perspective has proved useful in a broad range of problems, including quantile treatment effects \citep{chernozhukov2005iv}, quantile factor models \citep{chen2021quantile}, quantile autoregression \citep{koenker2006quantile}, and Granger causality in quantiles \citep{jeong2012consistent}. Nevertheless, extending the invariance principle from the conditional mean to the quantile framework is fundamentally non-trivial. In particular, the quantile score involves a nonlinear indicator transformation, and the corresponding first-order condition yields only an unconditional moment restriction that does not generally imply the corresponding conditional-quantile invariance, potentially giving rise to spurious invariant solutions. Moreover, the curvature of the quantile risk is local and depends on the conditional density at the target quantile, while the non-smooth empirical quantile loss requires additional treatment and, in our formulation, smoothing introduces an approximation bias. These features pose substantial challenges to the optimization, identification, and estimation of invariant conditional quantiles. The present work aims to fill this gap and capture quantile-dependent invariance that mean-based methods are unable to retrieve. Our analysis characterizes the role of environmental heterogeneity in identification, quantifies the effect of smoothing, and establishes non-asymptotic guarantees for estimation and variable selection. Before presenting our model, we first review related work including ``invariance'' under heterogeneous environments.

\vspace{-0.2cm}

\subsection{Related Work}\label{sec1.1}
Statistical regression with data collected from multiple environments is a prevalent paradigm in modern applications \citep{meinshausen2016methods,vcuklina2021diagnostics}. A line of work is the so-called transfer learning, with examples including nonparametric classification \citep{cai2021transfer}, transfer reinforcement learning \citep{chai2025deep}, and parametric or semiparametric regression models \citep{li2022transfer,bastani2021predicting,tian2023transfer,zhang2025transfer,jin2024transfer,huang2022estimation,fu2025transfer,zhang2024prediction}. However, these works typically assume certain similarities across environments; most frequently, such similarity is quantified by the distance between the underlying true parameters. Such assumptions may be restrictive from the perspective of robust generalization. For example, classical quantile regression under transfer learning framework \citep{zhang2025transfer,jin2024transfer,huang2022estimation} may suffer from overfitting or reduced performance due to spurious environment-dependent correlations. This perspective fundamentally differs from the invariance-based causal inference. A prominent approach to invariance-based causal inference is the ``Invariant Causal Prediction'' (ICP) framework introduced in the seminal work of \citet{peters2016causal}, based on the linear model: $y^{(e)}=\boldsymbol{\beta}_{S^*}^{*\top}\boldsymbol{x}_{S^*}^{(e)}+\varepsilon^{(e)}~\text{with}~\varepsilon^{(e)}\perp\!\!\!\perp\boldsymbol{x}_{S^*}^{(e)}~\text{and}~\mathbb{E}(\varepsilon^{(e)})=0,~\forall e\in\mathcal{E}$, where $\mathcal{E}$ is the set of environments and $n^{(e)}$ observations $\{(\boldsymbol{x}_i^{(e)},y_i^{(e)})\}_{i=1}^{n^{(e)}}\subseteq\mathbb{R}^{p}\times \mathbb{R}$ are independent and identically distributed (i.i.d.) drawn from the population $(\boldsymbol{x}^{(e)},y^{(e)})\sim \mu^{(e)}$ for each $e\in\mathcal{E}$. Here $\boldsymbol{\beta}^*$ is the true parameter vector, which is assumed to be invariant across environments, and $S^*=\mathrm{supp}(\boldsymbol{\beta}^*)$ is the index set of the truly important variables. They propose a hypothesis testing procedure that satisfies $\mathbb{P}(\hat{S}\subseteq S^*)\geq 1-\alpha$ for some confidence level $\alpha$, where $\hat{S}$ is their selected set via ICP. However, it provides no assurance on the power of the resulting tests. Moreover, such procedures can be overly conservative, including the extreme case $\hat{S}=\emptyset$. In nonlinear ICP, \citet{heinze2018invariant} use invariant conditional quantile prediction to test whether exceedances of pooled quantile fits depend on the environment. Their procedure tests candidate variable sets for invariance, whereas our goal is to estimate a common linear coefficient at a specified quantile.

Beyond the ICP method, a substantial line of work grounded in the invariance principle has developed sample-efficient regression methodologies for estimating the causal parameter $\boldsymbol{\beta}^*$ \citep{ghassami2017learning,rothenhausler2019causal,rothenhausler2021anchor}. These methods, however, typically impose additional and often restrictive structural assumptions that substantially reduce the complexity of the original problem. On the optimization side, \citet{wang2024causal} propose the negative weight distributionally robust optimization (NegDRO) for risk-invariance. However, its theoretical guarantees rely on relatively strong assumptions on the additive intervention mechanism, which may also limit its applicability in more general settings. From the standpoint of out-of-distribution generalization, Invariant Risk Minimization (IRM; \citealp{arjovsky2019invariant}) aims to learn data representations for which the optimal variables are invariant across environments. Nonetheless, its requirement that $|\mathcal{E}|\geq p$ substantially limits its applicability in realistic settings.

To attain provably sample-efficient estimation of $\boldsymbol{\beta}^*$ and $S^*$ with a minimal identification condition, \citet{fan2024environment} propose an Environment Invariant Linear Least Squares (EILLS) estimator and establish its variable selection property and $\ell_2$ error bound under the model $y^{(e)}=\boldsymbol{\beta}_{S^*}^{*\top}\boldsymbol{x}_{S^*}^{(e)}+\varepsilon^{(e)}$ with $\mathbb{E}[\varepsilon^{(e)}|\boldsymbol{x}_{S^*}^{(e)}]\equiv0$, $\forall e\in\mathcal{E}$. The variable index set $[p]$ is then partitioned into three disjoint groups: (a) the truly important set $S^*$; (b) $G=\{j\notin S^*:\mathrm{Cov}^{\mathcal{E}}(\varepsilon^{(e)},\boldsymbol{x}_j^{(e)})\neq0\}$; and (c) $[p]\setminus(S^*\cup G)$. Here, $\mathrm{Cov}^{\mathcal{E}}(\varepsilon^{(e)},\boldsymbol{x}_j^{(e)})=|\mathcal{E}|^{-1}\sum_{e\in\mathcal{E}}\mathbb{E}[\varepsilon^{(e)}\boldsymbol{x}_j^{(e)}]$ is the pooled covariance between the noise and the $j$-th variable. Variables in group $G$ are termed endogenous variables, as they exhibit nonzero covariance with the noise term. Although including such variables may help predict the noise, they are irrelevant to the primary goal of causal discovery. Variables in the last group are exogenous variables, whose inclusion does not enhance estimation accuracy and contributes only marginal variance. The presence of $G$ induces the so-called ``curse of endogeneity'', which violates the key assumptions underlying most traditional least-squares-based approaches.

To attain a computationally, statistically efficient, and distributionally robust estimation, \citet{gu2025fundamental} propose a stronger identification condition than \citet{fan2024environment} to attain an interpolating estimator. Moreover, to break the barrier of linear structure, \citet{gu2025causality} consider the following model: $y^{(e)}=m^*(\boldsymbol{x}_{S^*}^{(e)})+\varepsilon^{(e)}~\text{with}~ \mathbb{E}[\varepsilon^{(e)}|\boldsymbol{x}_{S^*}^{(e)}]\equiv0$, where $m^*$ is the regression function that allows neural networks and nonparametric regression. However, as mentioned before, restricting attention to the conditional mean constrains both methodological performance and the comprehensive understanding of causal relationships.\vspace{-0.2cm}

\subsection{Model Formulation and Our Contributions}
To bridge this gap and advance invariant methods within a quantile-level guaranteed framework, we start from the perspective of quantile regression and introduce the quantile regression with invariant structure, named Invariant Quantile Regression (IQR). For each environment $e\in\mathcal{E}$, assume that we observe $n^{(e)}$ i.i.d. observations $\{(\boldsymbol{x}_i^{(e)},y_i^{(e)})\}_{i=1}^{n^{(e)}}$ drawn from the population $(\boldsymbol{x}^{(e)},y^{(e)})\sim \mu^{(e)}$. We consider the following IQR model:\vspace{-0.05cm}
\begin{align}\label{eq1.2}
    y^{(e)}=\boldsymbol{\beta}_{S^*}^{*\top}\boldsymbol{x}_{S^*}^{(e)}+\varepsilon^{(e)}\quad\text{with}\quad \mathbb{P}(\varepsilon^{(e)}\leq0|\boldsymbol{x}_{S^*}^{(e)})=\tau, \quad \forall e\in\mathcal{E},
\end{align}
where $\tau\in(0,1)$ is the quantile level, $\boldsymbol{\beta}^*\in\mathbb{R}^{p}$ is the true environment-invariant parameter under IQR and $S^*=\mathrm{supp}(\boldsymbol{\beta}^*)$ indicates its support. The objective is to recover $S^*$ and $\boldsymbol{\beta}^*$ for a given $\tau$ from the data $\{(\boldsymbol{x}_i^{(e)},y_i^{(e)})\}_{i\in[n^{(e)}],e\in\mathcal{E}}$. From model \eqref{eq1.2}, the endogenous variables are defined as those indexed by 
$G=\big\{j\notin S^*:\sum_{e\in\mathcal{E}}\mathbb{E}[\{\mathds{1}(\varepsilon^{(e)}\leq0)-\tau\}\boldsymbol{x}_{j}^{(e)}]\neq 0\big\}$, where $\mathds{1}(\cdot)$ is the indicator function. This formulation departs from traditional quantile regression models, in which the condition $\mathbb{E}[\{\mathds{1}(\varepsilon^{(e)}\leq0)-\tau\}\boldsymbol{x}_{j}^{(e)}]=0$ for all $j$ arises as a first-order sub-gradient optimality condition. However, this assumption is violated in the presence of $G$, leading to the so-called ``curse of endogeneity'' where spurious correlations contribute to predicting noise but bias the estimates. Our IQR framework addresses this issue since it allows the existence of $G$. To estimate $\boldsymbol{\beta}^*$ and identify $S^*$ under Model \eqref{eq1.2}, for a fixed $\tau$, when we assign equal weights to each environment, we consider the following population-level loss function:\vspace{-0.1cm}
\begin{align*}
    \mathsf{Q}(\boldsymbol{\beta};\gamma)=\mathsf{R}(\boldsymbol{\beta})+\gamma\sum_{j=1}^{p}\mathds{1}(\beta_j\neq0)\sum_{e\in\mathcal{E}}\left|\mathbb{E}[(\mathds{1}(y^{(e)}-\boldsymbol{\beta}^\top \boldsymbol{x}^{(e)}\leq0)-\tau)\boldsymbol{x}_{j}^{(e)}]\right|^2,
\end{align*}
where $\mathsf{R}(\boldsymbol{\beta})=\sum_{e\in\mathcal{E}}\mathbb{E}\left[\rho_\tau(y^{(e)}-\boldsymbol{\beta}^\top\boldsymbol{x}^{(e)})\right]$ is the population-level pooled quantile loss function and $\rho_\tau(u)=u(\tau-\mathds{1}(u<0))$ is the check function. The introduction of the latter term can be viewed as the goal of invariance pursuit, while $\mathsf{R}(\boldsymbol{\beta})$ is employed to avoid trivial solutions such as $\boldsymbol{\beta}=\boldsymbol{0}_p$. In particular, the latter term encourages finding a solution $\hat{\boldsymbol{\beta}}$ such that 
$\mathbb{E}[(\mathds{1}(y^{(e)}-\hat{\boldsymbol{\beta}}^\top \boldsymbol{x}^{(e)}\leq0)-\tau)\boldsymbol{x}_{j}^{(e)}]=0,~\forall e\in\mathcal{E},~\forall j\in\mathrm{supp}(\hat{\boldsymbol{\beta}})$, which serves as a relaxation of $\mathbb{P}(y^{(e)}-\hat{\boldsymbol{\beta}}^\top \boldsymbol{x}^{(e)}\leq0|\boldsymbol{x}_{\mathrm{supp}(\hat{\boldsymbol{\beta}})})=\tau$ for any $e\in\mathcal{E}$. Furthermore, the quantile loss is non-smooth and non-differentiable, posing challenges for both computation and theoretical analysis. Motivated by recent advances in smoothed quantile regression (SQR) \citep{tan2022high,tan2022communication,xie2025statistical,zhang2025transfer,fernandes2021smoothing}, which yields a twice continuously differentiable and locally strongly convex function with high probability and is therefore computationally efficient, we consider the population loss alternatively: $\mathsf{Q}_{h}(\boldsymbol{\beta};\gamma)=\mathsf{R}_h(\boldsymbol{\beta})+\gamma\sum_{j=1}^{p}\mathds{1}(\beta_j\neq0)\sum_{e\in\mathcal{E}}\left|\nabla_j\mathsf{R}_h^{(e)}(\boldsymbol{\beta})\right|^2$, where $\mathsf{R}_h(\boldsymbol{\beta})$ can be regarded as a kernel-smoothed version of $\mathsf{R}(\boldsymbol{\beta})$. This smoothing technique facilitates both theoretical analysis and practical implementation of our approach. Our contribution is summarized as follows.\vspace{-0.3cm}
\begin{itemize}
    \item \textbf{Invariant Quantile Regression Framework.} We introduce an IQR framework that formalizes quantile regression under an invariance structure across heterogeneous environments and give the invariant parameter a worst-case quantile risk interpretation. Compared with existing invariant approaches that focus on the conditional mean, our framework captures the invariance across different quantile levels, thereby providing a more nuanced view of the data, including both the central behavior and the tail phenomenon, and enhances robustness to heavy-tailed noise. Moreover, unlike classical quantile regression under transfer learning framework, IQR isolates the invariant components of the conditional quantile function that persist across environments, thereby improving robustness and generalization.\vspace{-0.15cm}
    \item \textbf{Kernel-smoothed invariance-type estimator with theoretical guarantees.} We propose a Kernel-Smoothed Invariant Quantile Regression (KS-IQR) estimator and establish non-asymptotic guarantees for global minimizers over a bounded parameter space. Under an identification condition (Assumption \ref{assum4.5}) and suitable tuning conditions, the estimator excludes endogenous variables and retains the true support when a minimum signal condition holds. We further quantify the smoothing-induced deviation from exact invariance and establish a non-asymptotic $\ell_2$ error bound.\vspace{-0.15cm}
    \item \textbf{Enhanced sensitivity to causal discovery and endogenous variables.} As a byproduct of our theoretical results, the proposed estimator provides a more comprehensive characterization of the conditional distribution of the outcome, offering deeper insights into causal discovery. Moreover, compared with conditional-mean approaches, the IQR framework can help exclude spurious variables by detecting endogeneity at particular quantiles, as illustrated in Model~2 of Section \ref{sec5}.
\end{itemize}


\vspace{-0.2cm}

\noindent
\textbf{Notation:} For a vector $\boldsymbol{v}=(v_1,\ldots,v_p)^\top\in\mathbb{R}^{p}$, denote $\|\boldsymbol{v}\|_2=(\sum_{i=1}^{p}|v_i|^2)^{1/2}$, $\|\boldsymbol{v}\|_1=\sum_{i=1}^{p}|v_i|$, $\|\boldsymbol{v}\|_\infty=\max_{1\leq i\leq p}|v_i|$. Denote $[p]=\{1,\ldots,p\}$. We use $\mathrm{supp}(\boldsymbol{v})=\{j\in[p]:v_j\neq 0\}$ to indicate the support set of $\boldsymbol{v}$. We write $[\boldsymbol{v}]_{S}$ as the sub-vector of $\boldsymbol{v}$ for any $S\subseteq[p]$ and abbreviate it as $\boldsymbol{v}_{S}$ if without ambiguity. Let $\nabla_{S}f(\boldsymbol{v})$ be the gradient of $f(\boldsymbol{v})$ with respect to $\boldsymbol{v}_{S}$ for a given function $f$. Denote $\mathbb{S}^{l-1}$ as the $l$-dimensional unit sphere in $\mathbb{R}^{l}$. For a matrix $\boldsymbol{A}=(A_{i,j})_{1\leq i\leq k,1\leq j\leq l}\in \mathbb{R}^{k\times l}$, denote $\|\boldsymbol{A}\|_2=\max_{\boldsymbol{v}\in\mathbb{S}^{l-1}}\|\boldsymbol{A}\boldsymbol{v}\|_2$. Write $\|\boldsymbol{v}\|_{\boldsymbol{A}}=\|\boldsymbol{A}^{1/2}\boldsymbol{v}\|_2$ to denote the vector norm induced by $\boldsymbol{A}$. We use $\boldsymbol{A}_{S,T}=[A_{i,j}]_{i\in S,j\in T}$ to represent the sub-matrix of $\boldsymbol{A}$ and abbreviate it as $\boldsymbol{A}_{S}$ if $S=T$. Define $\mathds{1}(\cdot)$ as the indicator function. We use $\mu$-a.s. to denote almost surely with respect to a probability measure $\mu$. For two sequences of non-negative numbers $\{a_n\}$ and $\{b_n\}$, we write $a_n\lesssim b_n$ if there exists a constant $C>0$ independent of $n$ such that $a_n\leq C b_n$. Similarly, $a_n\gtrsim b_n$ means that $b_n\lesssim a_n$.\vspace{-0.2cm}
\section{Problem Setup and Quantile Invariance}\label{sec2}\vspace{-0.17cm}
\subsection{Problem Formulation and Robust Interpretation}\label{sec2.1}\vspace{-0.13cm}
We first provide a robust prediction interpretation
of the invariant parameter $\boldsymbol{\beta}^*$.
For constants $\sigma_{\varepsilon}>0$, $\sigma_x\geq1$, and $f_u\in(0,+\infty]$, define $\mathcal{U}_{\tau}(f_u):=\mathcal{U}_{\tau}(\boldsymbol{\beta}^*;\sigma_{\varepsilon},\sigma_x,f_u)$, indexed by $\tau\in(0,1)$, as the collection of distributions $\mu$ of $(\boldsymbol{x},y)$ satisfying (where $[\boldsymbol{x}]_1\equiv1$ represents the intercept):\vspace{-0.2cm}
\begin{align*}
    \quad\mathbb{P}_{\mu}(\varepsilon\leq0|\boldsymbol{x}_{S^*})=\tau,\quad \mathbb{E}_{\mu}\!\left[\rho_{\tau}(\varepsilon)|\boldsymbol{x}_{S^*}\right]\leq\sigma_{\varepsilon},\quad\mu-\text{a.s.},\quad\text{where}\quad \varepsilon=y-\boldsymbol{\beta}^{*\top}\boldsymbol{x},\\
    f_{\varepsilon|\boldsymbol{x}}(u)\leq f_u \quad\text{for all }u\in\mathbb{R},\quad \mu-\text{a.s.},\qquad \max_{j\in[p]}\mathbb{E}_{\mu}[x_j^2]\leq \sigma_x^2, \qquad [\boldsymbol{x}]_1\equiv1,
\end{align*}
where $\rho_\tau(u)=u(\tau-\mathds{1}(u<0))$ is the check function. When $f_u<+\infty$, $f_{\varepsilon|\boldsymbol{x}}$ denotes the conditional density function of $\varepsilon$ given $\boldsymbol{x}$; When $f_u=+\infty$, neither the existence of such a density nor any density upper bound is required. This class allows the joint distribution of $(\boldsymbol{x},y)$ to vary while preserving
the invariant conditional $\tau$-quantile relationship on $S^*$. Define the robust out-of-sample quantile risk as $\mathsf{R}_{\mathrm{oos},\tau}\big(\boldsymbol{\beta};\mathcal{U}_{\tau}(f_u)\big)=\sup_{\mu\in\mathcal{U}_{\tau}(f_u)}\mathbb{E}_{\mu}\left[\rho_{\tau}\big(y-\boldsymbol{\beta}^{\top}\boldsymbol{x}\big)\right]$.\vspace{-0.13cm}
The following Proposition \ref{prop: robust oos quantile risk} shows that the invariant parameter
$\boldsymbol{\beta}^*$ remains optimal against distribution shifts that preserve the
conditional $\tau$-quantile structure on $S^*$.\vspace{-0.14cm}
\begin{proposition}
\label{prop: robust oos quantile risk}
We have $\mathsf{R}_{\mathrm{oos},\tau}\big(\boldsymbol{\beta}^*;\mathcal{U}_{\tau}(\infty)\big)=\sigma_{\varepsilon}$ and, for any $\boldsymbol{\beta}\in\mathbb{R}^p$,\vspace{-0.15cm}
\begin{align*}
c_{\tau}\min\left\{\!\frac{\|\boldsymbol{\beta}-\boldsymbol{\beta}^*\|_2^2}{\sigma_{\varepsilon}},\|\boldsymbol{\beta}-\boldsymbol{\beta}^*\|_2\!\right\}\leq&\mathsf{R}_{\mathrm{oos},\tau}\big(\boldsymbol{\beta};\mathcal{U}_{\tau}(\infty)\big)\!-\!\mathsf{R}_{\mathrm{oos},\tau}\big(\boldsymbol{\beta}^*;\mathcal{U}_{\tau}(\infty)\big)\\
\leq&(\tau\vee(1-\tau))p^{1/2}\sigma_x\|\boldsymbol{\beta}-\boldsymbol{\beta}^*\|_2,~\text{where}~c_\tau=\tau(1-\tau)/96,
\end{align*}\vspace{-0.15cm}
which implies that $\boldsymbol{\beta}^*$ is the unique minimizer. Moreover, for every $f_u\in(0,\infty)$ such that $\mathcal{U}_{\tau}(f_u)\neq\varnothing$, we have $\mathsf{R}_{\mathrm{oos},\tau}\big(\boldsymbol{\beta};\mathcal{U}_{\tau}(f_u)\big)\!-\!\mathsf{R}_{\mathrm{oos},\tau}\big(\boldsymbol{\beta}^*;\mathcal{U}_{\tau}(f_u)\big)\leq(\tau\vee(1-\tau))p^{1/2}\sigma_x\|[\boldsymbol{\beta}-\boldsymbol{\beta}^*]_{[p]\setminus S^*}\|_2+f_u p\sigma_x^2\|\boldsymbol{\beta}-\boldsymbol{\beta}^*\|_2^2/2$.
\end{proposition}
\vspace{-0.13cm}
In particular, when $f_{\varepsilon|\boldsymbol{x}}$ is upper bounded and $\mathrm{supp}(\boldsymbol{\beta})\subseteq S^*$, the excess robust quantile risk is at most the order of $\|\boldsymbol{\beta}-\boldsymbol{\beta}^*\|_2^2$. This motivates us to use $\|\boldsymbol{\beta}-\boldsymbol{\beta}^*\|_2$ as a generalization metric and estimate the true support set $S^*$ and the true parameter $\boldsymbol{\beta}^*$.\vspace{-0.08cm}

Now, for each $e\in\mathcal{E}$, assume that we observe $n^{(e)}$ i.i.d. samples $\{(\boldsymbol{x}_i^{(e)},y_i^{(e)})\}_{i=1}^{n^{(e)}}\subseteq\mathbb{R}^{p}\times\mathbb{R}$, drawn from a population $\mu^{(e)}$. Here, we always denote $[\boldsymbol{x}_{i}^{(e)}]_{1}\equiv1$ to represent the intercept. Recall our IQR model \eqref{eq1.2}. The dependence of $\boldsymbol{\beta}^*$, $S^*$, and $\varepsilon^{(e)}$ on $\tau$ will be assumed implicitly and omitted from the notation for brevity. Our goal is to estimate the true parameter $\boldsymbol{\beta}^*$ and identify its support set $S^*$. Formally, $S^*$ satisfies the $\tau$-CP-invariant property, defined as follows.\vspace{-0.14cm}
\begin{definition}[$\tau$-CP-invariant Property]\label{def2.1}
    A set $S\subseteq[p]$ is defined to have a \textbf{$\tau$-conditional probability invariant} ($\tau$-CP-invariant) property across environments $\mathcal{E}$ if there exists some $\boldsymbol{\beta}$ with $\mathrm{supp}(\boldsymbol{\beta})=S$ such that $\mathbb{P}(y^{(e)}-\boldsymbol{\beta}_{S}^\top\boldsymbol{x}_{S}^{(e)}\leq0|\boldsymbol{x}_{S}^{(e)})=\tau,~~\forall e\in\mathcal{E}$.
\end{definition}\vspace{-0.14cm}
Define $\mathbb{E}[f(\boldsymbol{x}^{(e)}, y^{(e)})] = \int f(\boldsymbol{x}, y) \mu^{(e)}(d\boldsymbol{x}, dy)$ for a measurable function $f$. Then we define the population quantile loss $\mathsf{R}^{(e)}(\boldsymbol{\beta})$ and empirical quantile loss $\hat{\mathsf{R}}^{(e)}(\boldsymbol{\beta})$ on the $e$-th environment as $\mathsf{R}^{(e)}(\boldsymbol{\beta}) = \mathbb{E} \left[\rho_\tau(y^{(e)} - \boldsymbol{\beta}^\top \boldsymbol{x}^{(e)})\right] ~ \text{and} ~ \hat{\mathsf{R}}^{(e)}(\boldsymbol{\beta}) = \sum_{i=1}^{n^{(e)}}\left[\rho_\tau(y_i^{(e)} - \boldsymbol{\beta}^\top \boldsymbol{x}_i^{(e)})\right]/n^{(e)}$. The pooled population quantile loss is defined as $\mathsf{R}(\boldsymbol{\beta})=\sum_{e\in\mathcal{E}}\omega^{(e)}\mathsf{R}^{(e)}(\boldsymbol{\beta})$, where $\boldsymbol{\omega}=(\omega^{(e)})_{e\in\mathcal{E}}$ is the pre-specified weight vector with $\omega^{(e)}>0$ and $\sum_{e\in\mathcal{E}}\omega^{(e)}=1$. Typical choices include $\omega^{(e)}=n^{(e)}/\sum_{e'\in\mathcal{E}}n^{(e')}$, $\omega^{(e)}=|\mathcal{E}|^{-1}$, or weights specified according to prior knowledge. 
For notational simplicity, let $\boldsymbol{x}^{(e)}\sim\mu_{\boldsymbol{x}}^{(e)}$ denote the marginal distribution of the covariates in environment $e$, and let $\mu_{\varepsilon|\boldsymbol{x}}^{(e)}(\cdot|\boldsymbol{x})$ denote the conditional distribution of $\varepsilon^{(e)}$ given $\boldsymbol{x}^{(e)}=\boldsymbol{x}$. For any $S\subseteq[p]$, we similarly use $\mu_{\boldsymbol{x}_S}^{(e)}$, $\mu_{\varepsilon|\boldsymbol{x}_S}^{(e)}(\cdot|\boldsymbol{x}_S)$, and $\mu_{\boldsymbol{x}_S,\varepsilon}^{(e)}$ to denote the corresponding marginal, conditional, and joint distributions. The associated environment-mixed measures, Radon--Nikodym derivatives, and change-of-measure identities are provided in Section A.1 of the Supplementary Material. For instance, $\bar{\mu}_{\boldsymbol{x}_{S}}=\sum_{e\in\mathcal{E}}\omega^{(e)}\mu_{\boldsymbol{x}_{S}}^{(e)}$ denotes the environment-mixed marginal distribution of $\boldsymbol{x}_{S}$.\vspace{-0.3cm}

\subsection{Convolution-Type Smoothed Quantile Loss Function}\label{sec2.2}
Now we revisit the population quantile loss. Let $F_{\varepsilon^{(e)}|\boldsymbol{x}^{(e)}}$ be the conditional distribution of $\varepsilon^{(e)}$ given $\boldsymbol{x}^{(e)}$. The population quantile loss for each $e\in\mathcal{E}$ can be rewritten as:
\begin{align}\label{eq2.1}
    \mathsf{R}^{(e)}(\boldsymbol{\beta})=\mathbb{E}_{\mu_{\boldsymbol{x}}^{(e)}}\left\{\int_{-\infty}^{\infty}\rho_\tau(u-\langle\boldsymbol{x}^{(e)},\boldsymbol{\beta}-\boldsymbol{\beta}^*\rangle)\mathrm{d}F_{\varepsilon^{(e)}|\boldsymbol{x}^{(e)}}(u)\right\},
\end{align}
where $\langle\cdot,\cdot\rangle$ denotes the inner product, and $\boldsymbol{\beta}^*$ is the true invariant parameter, which coincides with the classical conditional quantile regression parameter when the conditional quantile restriction holds given the full covariate vector $\boldsymbol{x}^{(e)}$. Hence, $\mathsf{R}^{(e)}(\boldsymbol{\beta})$ is twice differentiable and strongly convex in a neighborhood of $\boldsymbol{\beta}^*$ provided sufficient smoothness of $F_{\varepsilon^{(e)}|\boldsymbol{x}^{(e)}}$. By replacing the unknown $F_{\varepsilon^{(e)}|\boldsymbol{x}^{(e)}}$ with the empirical cumulative distribution function (CDF) of the residuals $\{r_i^{(e)}(\boldsymbol{\beta}):=y_i^{(e)}-\boldsymbol{x}_i^{(e)\top}\boldsymbol{\beta}\}_{i=1}^{n^{(e)}}$, the empirical quantile loss can then be expressed as $\hat{\mathsf{R}}^{(e)}(\boldsymbol{\beta})=\int_{-\infty}^{\infty} \rho_\tau(u)\mathrm{d}\hat{F}^{(e)}(u;\boldsymbol{\beta}),$ where $\hat{F}^{(e)}(u;\boldsymbol{\beta})=\sum_{i=1}^{n^{(e)}}\mathds{1}(r_i^{(e)}(\boldsymbol{\beta})\leq u)/n^{(e)}$. However, $\hat{F}^{(e)}(u;\boldsymbol{\beta})$ is not continuous, which poses challenges for both computation and theoretical analysis of the minima of $\hat{\mathsf{R}}^{(e)}(\boldsymbol{\beta})$. Inspired by the SQR developed in \citep{fernandes2021smoothing,tan2022high}, we use the kernel CDF estimator. Define a sample size-dependent bandwidth parameter $h>0$, let $\hat{F}_h^{(e)}(\cdot;\boldsymbol{\beta})$ denote the smoothed distribution function corresponding to the Rosenblatt–Parzen kernel density estimator: 
$\hat{F}_h^{(e)}(u;\boldsymbol{\beta}) = \int_{-\infty}^u \hat{f}_h^{(e)}(t;\boldsymbol{\beta})\mathrm{d}t~\text{with}~
\hat{f}_h^{(e)}(t;\boldsymbol{\beta}) = \sum_{i=1}^{n^{(e)}}K_h(t-r_i^{(e)}(\boldsymbol{\beta}))/n^{(e)}$, where $K:\mathbb{R}\to[0,\infty)$ is a symmetric, non-negative kernel function satisfying $\int_{-\infty}^{\infty}K(u)\mathrm{d}u=1$, and 
$K_h(u):=K(u/h)/h,$ for $u\in\mathbb{R}$. Replacing $F_{\varepsilon^{(e)}|\boldsymbol{x}^{(e)}}$ in the population quantile loss with $\hat{F}_h^{(e)}(\cdot;\boldsymbol{\beta})$ leads to the following kernel-smoothed empirical quantile loss:\vspace{-0.18cm}
\begin{align*}
\hat{\mathsf{R}}_h^{(e)}(\boldsymbol{\beta})
=\int_{-\infty}^{\infty}\rho_\tau(u)\mathrm{d}\hat{F}_{h}^{(e)}(u;\boldsymbol{\beta}) 
=\frac{1}{n^{(e)}h}\sum_{i=1}^{n^{(e)}}\int_{-\infty}^{\infty}\rho_\tau(u)K\left(\frac{u+\boldsymbol{x}_i^{(e)\top}\boldsymbol{\beta}-y_i^{(e)}}{h}\right)\mathrm{d}u.
\end{align*}\vspace{-0.18cm}
Such a kernel-smoothed empirical quantile loss can be rewritten by means of the convolution operator: $\hat{\mathsf{R}}_h^{(e)}(\boldsymbol{\beta})=\!
\!\sum_{i=1}^{n^{(e)}}\ell_{h,\tau}(y_i^{(e)}\!\!-\boldsymbol{x}_i^{(e)\top}\!\!\boldsymbol{\beta})/n^{(e)}$, where $\ell_{h,\tau}(v):=(\rho_\tau\!*\!K_h)(v)\!=
\int_{-\infty}^{\infty}\rho_\tau(u)K_h(u\\-v)\mathrm{d}u$. The loss $\ell_{h,\tau}$ is referred as the smoothed quantile loss. For explicit expressions of $\ell_{h,\tau}$ under specific kernel choices, see Remark 2.1 in \citet{tan2022high}. Write $\bar{K}(v)=\int_{-\infty}^{v}K(u)\mathrm{d}u$ and $\bar{K}_h(v)=\int_{-\infty}^{v}K_h(u)\mathrm{d}u=\bar{K}(v/h)$. It can be verified that $\hat{\mathsf{R}}_h^{(e)}(\boldsymbol{\beta})$ is twice continuously differentiable with gradient $\nabla\hat{\mathsf{R}}_h^{(e)}(\boldsymbol{\beta})=(1/n^{(e)})\sum_{i=1}^{n^{(e)}}\{\bar{K}_{h}(-r_{i}^{(e)}\!(\boldsymbol{\beta}))-\tau\}\boldsymbol{x}_{i}^{(e)}$ and Hessian matrix
$\nabla^2\hat{\mathsf{R}}_h^{(e)}(\boldsymbol{\beta})=(1/n^{(e)})\sum_{i=1}^{n^{(e)}}K_{h}(-r_{i}^{(e)}(\boldsymbol{\beta}))\boldsymbol{x}_{i}^{(e)}\boldsymbol{x}_{i}^{(e)\top}$.\vspace{-0.1cm}

Define $\mathsf{R}_h^{(e)}(\boldsymbol{\beta})=\mathbb{E}[\hat{\mathsf{R}}_h^{(e)}(\boldsymbol{\beta})]$\footnote{If the loss is not integrable, we interpret $\mathsf{R}_h^{(e)}(\boldsymbol{\beta})$ as $\mathbb{E}[\hat{\mathsf{R}}_h^{(e)}(\boldsymbol{\beta})-\hat{\mathsf{R}}_h^{(e)}(\boldsymbol{0})]$, and use the same convention for $\mathsf{R}^{(e)}$ \citep{fernandes2021smoothing}.} and 
$\boldsymbol{\beta}_{h}^{*(e)}=\mathop{\arg\min}\limits_{\mathrm{supp}(\boldsymbol{\beta})\subseteq S^*}\mathsf{R}_{h}^{(e)}(\boldsymbol{\beta})$, it can be verified that, under mild conditions, $\|\boldsymbol{\beta}_{h}^{*(e)}-\boldsymbol{\beta}^*\|_2\lesssim h^2$, referred to as the smoothing bias \citep{tan2022high}, which is negligible for small bandwidth $h$. In the next section, we will construct the loss function for causal pursuit based on the smoothed quantile loss.

\vspace{-0.35cm}

\section{Kernel-Smoothed Invariant Quantile Regression}\label{sec3}\vspace{-0.2cm}
Recall the $\tau$-CP-invariant property in Definition \ref{def2.1}. For any set $S$ satisfying this property and for $e\in\mathcal{E}$, we have $\mathbb{P}(y^{(e)}-\boldsymbol{\beta}_{S}^\top\boldsymbol{x}_{S}^{(e)}\leq0|\boldsymbol{x}_{S}^{(e)})=\tau\Longleftrightarrow\mathbb{E}[\mathds{1}(y^{(e)}-\boldsymbol{\beta}_{S}^\top\boldsymbol{x}_{S}^{(e)}\leq0)-\tau|\boldsymbol{x}_{S}^{(e)}]=0\Longleftrightarrow\mathbb{E}[(\mathds{1}(y^{(e)}-\boldsymbol{\beta}_{S}^\top\boldsymbol{x}_{S}^{(e)}\leq0)-\tau)f(\boldsymbol{x}_{S}^{(e)})]=0$ for any measurable function $f$. Restricting the function to the linear function class yields \vspace{-0.13cm}$$\mathbb{P}(y^{(e)}-\boldsymbol{\beta}_{S}^\top\boldsymbol{x}_{S}^{(e)}\leq0|\boldsymbol{x}_{S}^{(e)})=\tau\Rightarrow\mathbb{E}[(\mathds{1}(y^{(e)}-\boldsymbol{\beta}_{S}^\top\boldsymbol{x}_{S}^{(e)}\leq0)-\tau)\boldsymbol{x}_{j}^{(e)}]=0,\quad\forall j\in S.$$\vspace{-0.13cm}
Here ``$\Longleftrightarrow$'' means equivalence and we make a relaxation such that $\mathbb{E}[(\mathds{1}(y^{(e)}-\boldsymbol{\beta}_{S}^\top\boldsymbol{x}_{S}^{(e)}\leq0)-\tau)\boldsymbol{x}_{S}^{(e)}]=0$ is a necessary but not sufficient condition for the original goal $\mathbb{P}(y^{(e)}-\boldsymbol{\beta}_{S}^\top\boldsymbol{x}_{S}^{(e)}\leq0|\boldsymbol{x}_{S}^{(e)})=\tau$. This inspires us to propose a focused invariance-type estimator by minimizing the following loss function at the population level:\vspace{-0.1cm}
\begin{align}\label{eq3.1}
    \mathsf{Q}(\boldsymbol{\beta};\!\gamma,\!\boldsymbol{\omega})\!=\!\!\underbrace{\sum_{e\in\mathcal{E}}\omega^{(e)}\mathsf{R}^{(e)}(\boldsymbol{\beta})}_{\mathsf{R}(\boldsymbol{\beta};\boldsymbol{\omega})}+\gamma\underbrace{\!\sum_{j=1}^{p}\!\mathds{1}(\beta_j\neq0)\!\sum_{e\in\mathcal{E}}\!\omega^{(e)}\!\left|\mathbb{E}[(\mathds{1}(y^{(e)}\!-\!\boldsymbol{\beta}^\top \!\boldsymbol{x}^{(e)}\!\leq0)\!-\!\tau)\boldsymbol{x}_{j}^{(e)}]\right|^2}_{\mathsf{J}(\boldsymbol{\beta};\boldsymbol{\omega})},
\end{align}\vspace{-0.13cm}
where $\mathsf{R}(\boldsymbol{\beta};\boldsymbol{\omega})$ is the pooled population-level quantile loss function, and $\gamma>0$ is a hyper-parameter. While the population-level objective in \eqref{eq3.1} provides a characterization of invariance and causal structure, its direct empirical implementation is hindered by the non-differentiability of the quantile loss and the indicator-based invariance penalty. This lack of smoothness poses substantial challenges for both computation and theoretical analysis. 

To address these difficulties, we adopt the kernel smoothing strategy reviewed in Section~\ref{sec2.2} to handle the non-differentiability of the quantile loss. This technique enables practical implementation via either brute force combined with the L-BFGS-B approach \citep{byrd1995limited,zhu1997algorithm}, or the Gumbel approximation \citep{maddison2017concrete,jang2017categorical} for solution search to handle the indicator function in $\mathsf{J}(\boldsymbol{\beta};\boldsymbol{\omega})$; see Section B.1 of the Supplementary Material for details. The resulting KS-IQR estimator, denoted by $\hat{\boldsymbol{\beta}}_{h,\mathsf{Q}}$, is defined as the minimizer of the empirical smoothed loss function for a given bandwidth $h$: $\hat{\boldsymbol{\beta}}_{h,\mathsf{Q}}=\mathop{\arg\min}\limits_{\boldsymbol{\beta}\in\mathbb{R}^{p}}\hat{\mathsf{Q}}_{h}(\boldsymbol{\beta};\gamma,\boldsymbol{\omega})$, where\vspace{-0.2cm}
\begin{align}\label{eq3.2}
    \hat{\mathsf{Q}}_{h}(\boldsymbol{\beta};\gamma,\boldsymbol{\omega})=\underbrace{\sum_{e\in\mathcal{E}}\omega^{(e)}\hat{\mathsf{R}}_{h}^{(e)}(\boldsymbol{\beta})}_{\hat{\mathsf{R}}_{h}(\boldsymbol{\beta};\boldsymbol{\omega})}+\gamma\underbrace{\sum_{j=1}^{p}\mathds{1}(\beta_j\neq0)\sum_{e\in\mathcal{E}}\omega^{(e)}\left|\nabla_j\hat{\mathsf{R}}_h^{(e)}(\boldsymbol{\beta})\right|^2}_{\hat{\mathsf{J}}_{h}(\boldsymbol{\beta};\boldsymbol{\omega})},
\end{align}\vspace{-0.1cm}
with $\hat{\mathsf{R}}_{h}(\boldsymbol{\beta};\boldsymbol{\omega})$ being the pooled empirical-level smoothed quantile loss function. For small $p$, we enumerate candidate supports and fit the restricted smooth objective on each support using L-BFGS-B. We report the candidate with the smallest fitted criterion value among the enumerated supports. For larger $p$, we use the Gumbel approximation described in Section B.1. The corresponding population-level KS-IQR estimator is to minimize\vspace{-0.15cm}
\begin{align}\label{eq3.3}
    \mathsf{Q}_{h}(\boldsymbol{\beta};\gamma,\boldsymbol{\omega})=\underbrace{\sum_{e\in\mathcal{E}}\omega^{(e)}\mathsf{R}_{h}^{(e)}(\boldsymbol{\beta})}_{\mathsf{R}_{h}(\boldsymbol{\beta};\boldsymbol{\omega})}+\gamma\underbrace{\sum_{j=1}^{p}\mathds{1}(\beta_j\neq0)\sum_{e\in\mathcal{E}}\omega^{(e)}\left|\nabla_j\mathsf{R}_h^{(e)}(\boldsymbol{\beta})\right|^2}_{\mathsf{J}_{h}(\boldsymbol{\beta};\boldsymbol{\omega})},
\end{align}
where $\mathsf{R}_{h}^{(e)}(\boldsymbol{\beta})=\mathbb{E}[\hat{\mathsf{R}}_{h}^{(e)}(\boldsymbol{\beta})]$ and $\mathsf{R}_{h}(\boldsymbol{\beta};\boldsymbol{\omega})$ is the pooled population-level smoothed quantile loss function. Now we analyze the implications of $\mathsf{J}_{h}(\boldsymbol{\beta};\boldsymbol{\omega})$ attaining its global minimum. It can be seen that for any fixed $\hat{\boldsymbol{\beta}}$, \vspace{-0.13cm}
$$\mathsf{J}_{h}(\hat{\boldsymbol{\beta}};\boldsymbol{\omega})=0~\Leftrightarrow~\nabla_{\mathrm{supp}(\hat{\boldsymbol{\beta}})}\mathsf{R}_h^{(e)}(\hat{\boldsymbol{\beta}})=0~,\forall e\in\mathcal{E}~\Leftrightarrow~\hat{\boldsymbol{\beta}}\in\mathop{\arg\min}\limits_{\boldsymbol{\beta}:\,\mathrm{supp}(\boldsymbol{\beta})\subseteq\mathrm{supp}(\hat{\boldsymbol{\beta}})}\mathsf{R}_h^{(e)}(\boldsymbol{\beta}),~\forall e\in\mathcal{E}.$$
Thus, the support $\mathrm{supp}(\hat{\boldsymbol{\beta}})$ of a solution $\hat{\boldsymbol{\beta}}$ to the equation $\mathsf{J}_{h}(\boldsymbol{\beta};\boldsymbol{\omega})=0$ satisfies the $\tau$-SQR-invariant property, which is formally defined as follows.\vspace{-0.07cm}
\begin{definition}[$\tau$-SQR-invariant Property and $\tau$-QR-invariant Property]\label{def3.1}
    A set $S\subseteq[p]$ is defined to have a \textbf{$\tau$-smoothed quantile regression invariant} ($\tau$-SQR-invariant) property across environments $\mathcal{E}$ if there exists some $\boldsymbol{\beta}$ with $\mathrm{supp}(\boldsymbol{\beta})=S$ such that $\boldsymbol{\beta}\in\mathop{\arg\min}\limits_{\boldsymbol{\beta}':\,\mathrm{supp}(\boldsymbol{\beta}')\subseteq S}\mathsf{R}_h^{(e)}(\boldsymbol{\beta}')=\mathop{\arg\min}\limits_{\boldsymbol{\beta}':\,\mathrm{supp}(\boldsymbol{\beta}')\subseteq S}\mathbb{E}\left[\ell_{h,\tau}(y_i^{(e)}-\boldsymbol{x}_i^{(e)\top}\boldsymbol{\beta}')\right],~\forall e\in\mathcal{E}.$
    Meanwhile, a set $S\subseteq[p]$ is defined to have a \textbf{$\tau$-quantile regression invariant} ($\tau$-QR-invariant) property across environments $\mathcal{E}$ if there exists some $\boldsymbol{\beta}$ with $\mathrm{supp}(\boldsymbol{\beta})=S$ such~that $\boldsymbol{\beta}\in\mathop{\arg\min}\limits_{\boldsymbol{\beta}':\,\mathrm{supp}(\boldsymbol{\beta}')\subseteq S}\mathbb{E}\left[\rho_{\tau}(y_i^{(e)}-\boldsymbol{x}_i^{(e)\top}\boldsymbol{\beta}')\right],~\forall e\in\mathcal{E}.$
\end{definition}
The $\tau$-QR-invariant property is weaker than the $\tau$-CP-invariant property since the latter can imply the former, while the converse is generally false. However, the two notions coincide under certain conditions. In particular, suppose that $\mu_{\boldsymbol{x}_{S}^{(e)}}$ belongs to a complete family $\mathcal{H}^{(e)}$ and that the conditional distribution of $y^{(e)}$ given $\boldsymbol{x}_{S}^{(e)}$ remains unchanged as the marginal distribution $\mu_{\eta}^{(e)}$ varies over $\mathcal{H}^{(e)}$. Assume further that the $\tau$-QR-invariant condition holds with the same $\boldsymbol{\beta}$ for every $e\in\mathcal{E}$ and every $\mu_{\eta}^{(e)}\in\mathcal H^{(e)}$,  and that $\mathbb{P}(y^{(e)}=\boldsymbol{\beta}_S^\top
\boldsymbol{x}_S^{(e)})=0$ under each such law.
The first-order condition then gives $\mathbb{E}_{\boldsymbol{x}_{S}^{(e)}\sim\mu_{\eta}^{(e)}}[(\mathbb{P}(y^{(e)}-\boldsymbol{\beta}_{S}^\top\boldsymbol{x}_{S}^{(e)}\leq0|\boldsymbol{x}_{S}^{(e)})-\tau)\boldsymbol{x}_{S}^{(e)}]=\boldsymbol{0}_{|S|}, \forall \mu_{\eta}^{(e)}\in\mathcal{H}^{(e)}$. By completeness, we have $\mathbb{E}[(\mathds{1}(y^{(e)}-\boldsymbol{\beta}_{S}^\top\boldsymbol{x}_{S}^{(e)}\leq0)-\tau)|\boldsymbol{x}_{S}^{(e)}]\boldsymbol{x}_{S}^{(e)}=\boldsymbol{0}_{|S|}~~\mu_{\boldsymbol{x}_{S}^{(e)}}\text{-a.s.}$ if $\sup_{\mu_{\eta}^{(e)}\in\mathcal{H}^{(e)}}\mathbb{E}_{\boldsymbol{x}_{S}^{(e)}\sim\mu_{\eta}^{(e)}}[\|\boldsymbol{x}_{S}^{(e)}\|_{\infty}]<\infty$, from which we further have $\mathbb{E}[(\mathds{1}(y^{(e)}-\boldsymbol{\beta}_{S}^\top\boldsymbol{x}_{S}^{(e)}\leq0)-\tau)|\boldsymbol{x}_{S}^{(e)}]=0~~\mu_{\boldsymbol{x}_{S}^{(e)}}\text{-a.s.}$ if $\mu_{\boldsymbol{x}_{S}^{(e)}}(\{\boldsymbol{0}_{|S|}\})=0$. Thus, the $\tau$-QR-invariant condition imposed uniformly over a complete family is sufficient for the $\tau$-CP-invariant property. According to previous discussions, if $\tilde{\boldsymbol{\beta}}$ satisfying $\mathsf{J}_{h}(\tilde{\boldsymbol{\beta}};\boldsymbol{\omega})=0$ with support $\tilde{S}$, then $\tilde{S}$ has the $\tau$-SQR-invariant property. The $\tau$-SQR-invariant property is closely related to the $\tau$-QR-invariant property, with the former being defined through the smoothed quantile loss. However, the $\tau$-SQR-invariant property may not hold exactly for $S^*$ since the smoothing bias can differ across environments. We will later show that, under mild conditions, $S^*$ satisfies the $(\tau,\delta)$-nearly-SQR-invariant property defined in Definition \ref{def4.1} for some small $\delta$. Consequently, minimizing $\mathsf{J}_{h}(\boldsymbol{\beta};\boldsymbol{\omega})$ provides a smoothed approximation to the $\tau$-QR-invariant criterion, with the approximation error controlled by the bandwidth $h$.

As a possible extension, one may strengthen the invariance criterion by incorporating nonlinear transformations. Specifically, the penalty term $\mathsf{J}_{h}(\boldsymbol{\beta};\boldsymbol{\omega})$ in \eqref{eq3.3} can be replaced by $\mathsf{J}_h'(\boldsymbol{\beta};\boldsymbol{\omega})\!\!=\!\!\!\sum_{j=1}^{p}\!\mathds{1}(\beta_j\neq0)\!\!\sum_{e\in\mathcal{E}}\!\omega^{(e)}\!\bigg\{\left|\mathbb{E}[\{\bar{K}_{h}(-r_{i}^{(e)}\!(\boldsymbol{\beta}))\!-\!\tau\}\boldsymbol{x}_{j}^{(e)}\!]\right|^2\!\!\!+\!\!\left|\mathbb{E}[\{\bar{K}_{h}(-r_{i}^{(e)}\!(\boldsymbol{\beta}))\!-\!\tau\}f(\boldsymbol{x}_{j}^{(e)})]\right|^2\bigg\}$, where $f$ is a nonlinear function. More generally, one may consider aggregating multiple nonlinear transformations to further enhance the penalty. 

\vspace{-0.24cm}

\section{Theoretical Analysis}\label{sec4}\vspace{-0.2cm}
\subsection{Effect of Smoothing on Invariance: Nearly-SQR-Invariant Property}\vspace{-0.2cm}
In this section, we demonstrate that kernel smoothing has a negligible impact on the invariance property. We first assume some regularity conditions below.\vspace{-0.2cm}
\begin{assumption}\label{assum4.1}
$\{(\boldsymbol{x}_i^{(e)},y_i^{(e)})\}_{i=1}^{n^{(e)}}\subseteq\mathbb{R}^{p}\times\mathbb{R}$ are i.i.d. copies of $(\boldsymbol{x}^{(e)},y^{(e)})\sim\mu^{(e)}$ for each $e\in\mathcal{E}$. The data from different environments are also assumed to be independent. We set $[\boldsymbol{x}_i^{(e)}]_{1}\equiv1$ to represent the intercept.
\end{assumption}\vspace{-0.15cm}
\begin{assumption}\label{assum4.2}
Let $\boldsymbol{\Sigma}^{(e)}=\mathbb{E}[\boldsymbol{x}^{(e)}\boldsymbol{x}^{(e)\top}]$. Assume that its eigenvalues satisfy $0<\lambda_L\leq\lambda_{\min}(\boldsymbol{\Sigma}^{(e)})\leq1\leq\lambda_{\max}(\boldsymbol{\Sigma}^{(e)})\leq\lambda_U<\infty$, where $\lambda_L,\lambda_U>0$ are universal constants across environments. Moreover, there exists some constant $\sigma_x\in[1,\infty)$ such that
$$\mathbb{P}\left(\left|\boldsymbol{u}^\top(\boldsymbol{\Sigma}_S^{(e)})^{-1/2}\boldsymbol{x}_S^{(e)}\right|\geq t\right)\leq2\exp\left(-\frac{t^2}{2\sigma_x^2\|\boldsymbol{u}\|_2^2}\right),~~\forall e\in\mathcal{E},~S\subseteq[p],~\boldsymbol{u}\in\mathbb{R}^{|S|},~t>0.$$
\end{assumption}
\vspace{-0.15cm}
\begin{assumption}\label{assum4.3}
Assume that the kernel function $K:\mathbb{R}\mapsto [0,\infty)$ is symmetric around $0$, and satisfies $\int_{-\infty}^{\infty}K(u)\mathrm{d}u=1$. Moreover, assume that $k_1=\int_{-\infty}^{\infty}|u|K(u)\mathrm{d}u<\infty$, $k_2=\int_{-\infty}^{\infty}u^2K(u)\mathrm{d}u<\infty$, $k_l=\min_{|u|\leq 1}K(u)>0$, and $k_u=\max_{u\in\mathbb{R}}K(u)<\infty$.
\end{assumption}

Assumption \ref{assum4.1} and Assumption \ref{assum4.2} are standard in regression models \citep{fan2024environment,zhang2025transfer,gu2025fundamental}. Assumption \ref{assum4.3} imposes mild conditions on the kernel function, where the assumption on $k_l=\min_{|u|\leq 1}K(u)>0$ is just for notation convenience, see Remark 4.1 of \citet{tan2022high}.

For any $S\subseteq[p]$, let $f_{\varepsilon^{(e)}|\boldsymbol{x}_{S}^{(e)}}$ and $f_{\varepsilon^{(e)}|\boldsymbol{x}^{(e)}}$ be the conditional density function of $\varepsilon^{(e)}$ given $\boldsymbol{x}_{S}^{(e)}$ and given $\boldsymbol{x}^{(e)}$, respectively. Define $f_{\varepsilon|\boldsymbol{x}_{S}}=\sum_{e\in\mathcal{E}}\omega^{(e)}\rho_{\boldsymbol{x}_{S}}^{(e)}f_{\varepsilon^{(e)}|\boldsymbol{x}_{S}^{(e)}}$ and $f_{\varepsilon|\boldsymbol{x}}=\sum_{e\in\mathcal{E}}\omega^{(e)}\rho_{\boldsymbol{x}}^{(e)}f_{\varepsilon^{(e)}|\boldsymbol{x}^{(e)}}$ as the corresponding environment-averaged conditional densities. We also define $F_{\varepsilon|\boldsymbol{x}_{S}}(u)=\sum_{e\in\mathcal{E}}\omega^{(e)}\rho_{\boldsymbol{x}_{S}}^{(e)}F_{\varepsilon^{(e)}|\boldsymbol{x}_{S}^{(e)}}(u)$ as the environment-averaged cumulative conditional distribution function. 
With these preparations, we define the $(\tau,\delta)$-nearly-SQR-invariant property as follows.
\begin{definition}[$(\tau,\delta)$-nearly-SQR-invariant Property]\label{def4.1}
     A set $S\subseteq[p]$ is defined to have a \textbf{$(\tau,\delta)$-nearly smoothed quantile regression invariant} ($(\tau,\delta)$-nearly-SQR-invariant) property if there exist some $\boldsymbol{\beta}$ and $\{\boldsymbol{\beta}^{(e)}\}_{e\in\mathcal{E}}$ with $\mathrm{supp}(\boldsymbol{\beta})=S$ and $\mathrm{supp}(\boldsymbol{\beta}^{(e)})\subseteq S$ for all $e\in\mathcal{E}$ such that
     \begin{align*}
         &\boldsymbol{\beta}\in\mathop{\arg\min}\limits_{\boldsymbol{\beta}':\,\mathrm{supp}(\boldsymbol{\beta}')\subseteq S}\mathbb{E}\left[\rho_{\tau}(y_i^{(e)}-\boldsymbol{x}_i^{(e)\top}\boldsymbol{\beta}')\right],~\forall e\in\mathcal{E},\\
         &\boldsymbol{\beta}^{(e)}\in\mathop{\arg\min}\limits_{\boldsymbol{\beta}':\,\mathrm{supp}(\boldsymbol{\beta}')\subseteq S}\mathsf{R}_h^{(e)}(\boldsymbol{\beta}'),~\text{and}~\|\boldsymbol{\beta}_{S}-\boldsymbol{\beta}_{S}^{(e)}\|_{\boldsymbol{\Sigma}_{S}^{(e)}}\leq\delta,~\forall e\in\mathcal{E}.
     \end{align*}
\end{definition}

Now we argue that $S^*$ possesses the $(\tau,\delta)$-nearly-SQR-invariant property for a small $\delta$ when the bandwidth $h$ is chosen to be small.

\begin{proposition}[$S^*$: $(\tau,\delta)$-nearly-SQR-invariant property]\label{prop:nearly invariant set}
    Under Assumptions \ref{assum4.1}--\ref{assum4.3}, assume that $f_l\leq f_{\varepsilon^{(e)}|\boldsymbol{x}_{S^*}^{(e)}}(0)\leq f_u$ for some $f_u\geq f_l>0$ and $f_{\varepsilon^{(e)}|\boldsymbol{x}_{S^*}^{(e)}}(u)$ is $l_0$-Lipschitz in $u$ for $\mu_{\boldsymbol{x}_{S^*}}^{(e)}$-a.s. $\boldsymbol{x}_{S^*}^{(e)}$ and each $e\in\mathcal{E}$, that is, $|f_{\varepsilon^{(e)}|\boldsymbol{x}_{S^*}^{(e)}}(u)-f_{\varepsilon^{(e)}|\boldsymbol{x}_{S^*}^{(e)}}(v)|\leq l_0|u-v|$ for all $u,v\in\mathbb{R}$. Define $c_0=0.5\mu_3+k_1+0.5k_2$ and $\mu_3=3\sqrt{2\pi}\sigma_x^3$, then $S^*$ satisfies the $(\tau,(c_0l_0/f_l)h^2)$-nearly-SQR-invariant property provided that $h<f_l /(c_0l_0)$.
\end{proposition}
From Proposition \ref{prop:nearly invariant set}, the objective function $\mathsf{J}_{h}(\boldsymbol{\beta};\boldsymbol{\omega})$ is to seek an approximation of the $\tau$-CP-invariant property. Our estimator can be regarded as the trade-off between the pooled smoothed quantile loss function and the objective function $\mathsf{J}_{h}(\boldsymbol{\beta};\boldsymbol{\omega})$ with causal pursuit. When $\gamma$ is properly selected, the spurious variables can be expected to be screened out.

\vspace{-0.2cm}

\subsection{A Local Quadratic Lower Bound for the Population Objective}\vspace{-0.1cm}
We first partition the variable index set $[p]\setminus S^*$ into endogenous and exogenous spurious components, defined as follows.

\begin{definition}[$\tau$-Pooled Endogenously and Exogenously Spurious Variables]\label{def4.2}
    For given weights $\boldsymbol{\omega}=(\omega^{(e)})_{e\in\mathcal{E}}$ and $\tau\in(0,1)$, let $G_{\boldsymbol{\omega}}$ denote the index set of endogenously spurious variables in environments $\mathcal{E}$, defined as $G_{\boldsymbol{\omega}}=\{j\notin S^*:\sum_{e\in\mathcal{E}}\omega^{(e)}\mathbb{E}[\{\mathds{1}(\varepsilon^{(e)}\leq0)-\tau\}\boldsymbol{x}_{j}^{(e)}]\neq 0\}$. Concurrently, let $L_{\boldsymbol{\omega}}$ be the index set of exogenously spurious variables, given by $L_{\boldsymbol{\omega}}=[p]\setminus (G_{\boldsymbol{\omega}}\cup S^*)$. Furthermore, we define the environment-weighted endogeneity as $\iota_j=\sum_{e\in\mathcal{E}}\omega^{(e)}\mathbb{E}[\{\mathds{1}(\varepsilon^{(e)}\leq0)-\tau\}\boldsymbol{x}_{j}^{(e)}]$ for $j\in G_{\boldsymbol{\omega}}$ and denote the minimum signal strength of endogeneity as $\iota=\min_{j\in G_{\boldsymbol{\omega}}}|\iota_{j}|>0$.
\end{definition}
The notation $G_{\boldsymbol{\omega}}$ and $L_{\boldsymbol{\omega}}$ emphasizes their potential dependence on the weight $\boldsymbol{\omega}$. The dependence on $\tau$ will be assumed implicitly and omitted from the notation for brevity.
\begin{assumption}\label{assum4.4}
Assume that (i) $f_{\varepsilon|\boldsymbol{x}}$ satisfies $f_l\leq f_{\varepsilon|\boldsymbol{x}}(0)\leq f_u$ for $\bar{\mu}_{\boldsymbol{x}}$-a.s. $\boldsymbol{x}$ and some constants $0<f_l\leq f_u<\infty$; (ii) $f_{\varepsilon|\boldsymbol{x}}(u)$ is $l_0$-Lipschitz in $u$ for $\bar{\mu}_{\boldsymbol{x}}$-a.s. $\boldsymbol{x}$ and some bounded constant $l_0>0$; (iii) $f_{\varepsilon^{(e)}|\boldsymbol{x}_{S^*}^{(e)}}(u)$ is $l_0$-Lipschitz in $u$ for $\mu_{\boldsymbol{x}_{S^*}}^{(e)}$-a.s. $\boldsymbol{x}_{S^*}^{(e)}$, and each $e\in\mathcal{E}$; (iv) $f_{\varepsilon^{(e)}|\boldsymbol{x}^{(e)}}(u)\leq f_u$ for all $u\in\mathbb{R}$, $\mu_{\boldsymbol{x}}^{(e)}$-a.s. $\boldsymbol{x}$ and each $e\in\mathcal{E}$; and (v) $f_{\varepsilon|\boldsymbol{x}_{L_{\boldsymbol{\omega}}}}(u)$ and $f_{\varepsilon|\boldsymbol{x}_{G_{\boldsymbol{\omega}}}}(u)$ are both $l_0$-Lipschitz in $u$ for $\bar{\mu}_{\boldsymbol{x}_{L_{\boldsymbol{\omega}}}}$-a.s. $\boldsymbol{x}_{L_{\boldsymbol{\omega}}}$ and $\bar{\mu}_{\boldsymbol{x}_{G_{\boldsymbol{\omega}}}}$-a.s. $\boldsymbol{x}_{G_{\boldsymbol{\omega}}}$, respectively.
\end{assumption}

Assumption \ref{assum4.4} is mild and is to ensure that the smoothing bias in the first order condition is negligible at the population-level for $S\in\{G_{\boldsymbol{\omega}},L_{\boldsymbol{\omega}},S^*\}$, i.e., \vspace{-0.1cm}$$\sum_{e\in\mathcal{E}}\omega^{(e)}\mathbb{E}[\{\mathds{1}(\varepsilon^{(e)}\leq0)-\tau\}\boldsymbol{x}_{S}^{(e)}]-\sum_{e\in\mathcal{E}}\omega^{(e)}\mathbb{E}[\bar{K}_{h}(-\varepsilon^{(e)})-\tau\}\boldsymbol{x}_{S}^{(e)}]$$
is small enough in some sense. Now, we show a quadratic lower bound for the population loss around $\boldsymbol{\beta}^*$ when $\gamma$ exceeds a certain critical threshold, which is crucial to establish the convergence towards $\boldsymbol{\beta}^*$. For $r>0$, define the rescaled $\ell_2$ ball as $\mathbb{B}_{\boldsymbol{\Sigma}}(r)=\left\{\boldsymbol{\delta}\in\mathbb{R}^{p}:\|\boldsymbol{\delta}\|_{\boldsymbol{\Sigma}}\leq r\right\}$, where $\boldsymbol{\Sigma}=\sum_{e\in\mathcal{E}}\omega^{(e)}\boldsymbol{\Sigma}^{(e)}$ is the pooled population-level covariance matrix. \vspace{-0.07cm}
\begin{theorem}\label{thm: locally strongly convex}
    Suppose that Assumptions \ref{assum4.1}--\ref{assum4.4} hold. For any $\boldsymbol{\beta}\in\mathbb{R}^{p}$ with $\boldsymbol{\beta}-\boldsymbol{\beta}^*\in\mathbb{B}_{\boldsymbol{\Sigma}}(r)$ and $r=h/(16\sqrt{2}\sigma_x^2\lambda_U/\lambda_L)$, provided that 
    $$h\leq\min\left\{\frac{f_l}{2l_0},\frac{4\sqrt{2}\iota\lambda_U\sigma_x^2}{\lambda_L^{1/2}\sqrt{1+\lambda_U^2f_u^2}},\left(\frac{\iota^2}{4(1+\lambda_U^2f_u^2)l_0^2k_2^2\lambda_{U}}\right)^{1/4}\right\}$$
    and $\gamma\geq\gamma_0^*:=16\lambda_U(1+\lambda_U^2f_u^2)/(\iota^2f_lk_l\lambda_L^2)\sup_{S\subseteq[p]:\,S\cap G_{\boldsymbol{\omega}}\neq\emptyset}\mathsf{b}_{S}$, we have \vspace{-0.07cm}
\begin{align*}
    \mathsf{Q}_h(\boldsymbol{\beta};\gamma,\boldsymbol{\omega})-\mathsf{Q}_h(\boldsymbol{\beta}^*;\gamma,\boldsymbol{\omega})\geq&\frac{13}{64}k_lf_l\lambda_L\|\boldsymbol{\beta}-\boldsymbol{\beta}^*\|_2^2+\left(\frac{\gamma\iota^2}{8(1+\lambda_U^2f_u^2)}-\frac{2\lambda_U}{f_lk_l\lambda_L^2}\mathsf{b}_{S}\right)-\frac{l_0^2k_2^2\lambda_U^2}{2f_lk_l\lambda_L^2}h^4\\
    \geq&\frac{13}{64}k_lf_l\lambda_L\|\boldsymbol{\beta}-\boldsymbol{\beta}^*\|_2^2-\frac{l_0^2k_2^2\lambda_U^2}{2f_lk_l\lambda_L^2}h^4~~~~~~~\text{if}~~\mathrm{supp}(\boldsymbol{\beta})\cap G_{\boldsymbol{\omega}}\neq\emptyset,
\end{align*}
where $\mathsf{b}_{S}=\left\|\sum_{e\in\mathcal{E}}\omega^{(e)}\mathbb{E}\left[\left(\bar{K}_{h}(-\varepsilon^{(e)})-\tau\right)\boldsymbol{x}_{S}^{(e)}\right]\right\|_2^2$, and if $\mathrm{supp}(\boldsymbol{\beta})\cap G_{\boldsymbol{\omega}}=\emptyset$, we have\vspace{-0.07cm}
\begin{align*}
    \mathsf{Q}_h(\boldsymbol{\beta};\gamma,\boldsymbol{\omega})-\mathsf{Q}_h(\boldsymbol{\beta}^*;\gamma,\boldsymbol{\omega})\geq\frac{13}{64}k_lf_l\lambda_L\|\boldsymbol{\beta}-\boldsymbol{\beta}^*\|_2^2-\frac{l_0^2k_2^2\lambda_U^2}{f_lk_l\lambda_L^2}h^4-\gamma\frac{l_0^2k_2^2\lambda_{U}}{4}h^4,
\end{align*}
\end{theorem}\vspace{-0.1cm}

Theorem \ref{thm: locally strongly convex} gives a local quadratic lower bound for the population objective around $\boldsymbol{\beta}^*$, up to an error of order $(1\vee\gamma)h^4$, provided that $\gamma\geq\gamma_0^*$ and $\boldsymbol{\beta}-\boldsymbol{\beta}^*\in\mathbb{B}_{\boldsymbol{\Sigma}}(r)$. Here and below, if $G_{\boldsymbol{\omega}}=\emptyset$, we omit the conditions involving $\iota$ or $s_-$ and set $\gamma_0^*=\gamma^*=0$. We use this lower bound in the finite-sample analysis below.
\vspace{-0.2cm}
\subsection{Non-asymptotic Estimation and Variable Selection of the KS-IQR Estimator}
Theorem \ref{thm: locally strongly convex} gives a local quadratic lower bound for the population objective around $\boldsymbol{\beta}^*$. For the finite-sample analysis, we impose the following identification condition, motivated by \citet{fan2024environment}: the population penalty must be bounded away from zero over all feasible parameters whose supports intersect $G_{\boldsymbol{\omega}}$. Let $\Theta_C=\{\boldsymbol{\beta}:~\|\boldsymbol{\beta}\|_2\leq C\}$.\vspace{-0.17cm}
\begin{assumption}[Identification Condition]\label{assum4.5}
Assume that $\boldsymbol{\beta}^*\in\Theta_{C_{n,p}}$ for some $C_{n,p}$, and $s_-:=\inf_{\boldsymbol{\beta}:\mathrm{supp}(\boldsymbol{\beta})\cap G_{\boldsymbol{\omega}}\neq\emptyset,\boldsymbol{\beta}\in\Theta_{C_{n,p}}}\mathsf{J}_h(\boldsymbol{\beta};\boldsymbol{\omega})>0$.
\end{assumption}\vspace{-0.07cm}

For the finite-sample analysis, we consider the constrained estimator:\vspace{-0.16cm}
\begin{align}
    \hat{\boldsymbol{\beta}}_{h,\mathsf{Q}}\in\mathop{\arg\min}_{\boldsymbol{\beta}\in\Theta_{C_{n,p}}}\hat{\mathsf{Q}}_{h}(\boldsymbol{\beta};\gamma,\boldsymbol{\omega}).
\end{align}
Throughout this subsection, $\hat{\boldsymbol{\beta}}_{h,\mathsf{Q}}$ denotes this estimator. Its existence follows from the compactness of $\Theta_{C_{n,p}}$ and the lower semicontinuity of $\hat{\mathsf{Q}}_{h}$. This estimator agrees with the unconstrained formulation in \eqref{eq3.2} whenever an unconstrained global minimizer belongs to $\Theta_{C_{n,p}}$. We allow $C_{n,p}$ to diverge with the sample sizes and dimension. 

\begin{remark}\label{remark 4.1}
    Assumption \ref{assum4.5} serves as an identifiability condition for the proposed score-based invariance criterion. Indeed, if there exists some $\boldsymbol{\beta}$ with $\mathrm{supp}(\boldsymbol{\beta})=S$ and $S\cap G_{\boldsymbol{\omega}}\neq\emptyset$ such that $\nabla_{S}\mathsf{R}^{(e)}(\boldsymbol{\beta})=\boldsymbol{0}_{|S|}$ for all $e\in\mathcal{E}$, then $S$ also satisfies the $\tau$-QR-invariant property. This implies that $S$ also satisfies the $\tau$-CP invariant property under the additional completeness and uniform-in-family $\tau$-QR-invariant conditions discussed in Section \ref{sec3}. In this situation, distinguishing between $S$ and $S^*$ becomes impossible since both satisfy the $\tau$-CP-invariant property, yet $S\setminus S^*\neq\emptyset$. Assumption \ref{assum4.5} excludes such competing solutions under the smoothed criterion by imposing $\mathsf{J}_h(\boldsymbol{\beta};\boldsymbol{\omega})\geq s_->0$ uniformly over $\boldsymbol{\beta}\in\Theta_{C_{n,p}}$ with $\mathrm{supp}(\boldsymbol{\beta})\cap G_{\boldsymbol{\omega}}\neq\emptyset$. With Gaussian kernel at a fixed bandwidth $h>0$, Model~3 in Section~5 satisfies this assumption on any fixed parameter ball containing $\boldsymbol{\beta}^*$ whenever $a_1\ne a_2$.
\end{remark}

Now we are ready to provide the theoretical properties of our constrained KS-IQR estimator. Assume that $\omega^{(e)}\equiv1/|\mathcal{E}|$ and $n^{(e)}\equiv n$ for each $e\in\mathcal{E}$ for notation simplicity. Without loss of generality, we also assume that $\log(|\mathcal{E}|)\leq c p$ for some constant $c<\infty$. Further details regarding varying $(n^{(e)},\omega^{(e)})$ can be found in Section A.4 of the Supplementary Material.
\begin{theorem}[Non-asymptotic $\ell_2$ Error Bound]\label{thm:non-asymptotic ell_2 Error Bound}
    Suppose that Assumptions \ref{assum4.1}--\ref{assum4.5} hold. Let $\xi_t\!=\!p(\log C_{n,p}\!\vee\!\log (n|\mathcal{E}|)\!\vee \!\log p)+t$. If $n\gtrsim \!\gamma\xi_t\!\vee\!\gamma^2(|{S^*}|+t+\log|\mathcal{E}|+1)$, $h\leq f_l/(2l_0)$, \vspace{-0.2cm}
    \begin{align*}
    &\sqrt{\frac{\xi_t}{n|\mathcal{E}|}\gamma}\vee\frac{p\log p}{n|\mathcal{E}|}\vee\gamma\sqrt{|S^*|}\left(\frac{t+\log(2|\mathcal{E}||{S^*}|)}{n\min_{j\in S^*}|\beta^*_j|}\right)\lesssim h\lesssim\gamma^{-1/2}\wedge\frac{n}{(\xi_t+\log(2|\mathcal{E}|))\gamma^2},\\
    &h\gtrsim \gamma^{2/3}\left(\frac{\xi_t(|S^*|+t+\log(2|\mathcal{E}|))}{n^{2}}\right)^{1/3}\vee\gamma^{1/2}\left(\frac{\xi_t^2(|S^*|+t+\log(2|\mathcal{E}|))}{n^{3}}\right)^{1/4},\\
    &cs_-\geq\frac{\xi_t}{n|\mathcal{E}|}+\frac{|S^*|+t+1+\log(|\mathcal{E}|)}{n}+h^4
\end{align*}\vspace{-0.1cm}
    for some sufficiently small constant $c>0$, and $\gamma\geq\! \gamma^*\vee1\!$ with \vspace{-0.1cm} $$\gamma^*=\frac{4\lambda_U}{f_lk_l\lambda_L^2}\!\sup_{S\subseteq[p]:\,S\cap G_{\boldsymbol{\omega}}\neq\emptyset}\frac{\mathsf{b}_{S}}{\inf_{\substack{\boldsymbol{\beta}\in\Theta_{C_{n,p}}\\\mathrm{supp}(\boldsymbol{\beta})=S}}\mathsf{J}_{h}(\boldsymbol{\beta};\boldsymbol{\omega})}\!\vee\! 16\sqrt{2}\lambda_U^{1/2}\!\left[1+C_0'\sigma_x^2\!\left(\!\sqrt{\frac{p+t}{n|\mathcal{E}|}}\!+\!\frac{p+t}{n|\mathcal{E}|}\!\right)\!\right]^{1/2}\!\!C_{n,p}/s_-,$$
    we obtain that, with probability at least $1-10e^{-t}$, \vspace{-0.07cm}
    \begin{align*}
        \|\hat{\boldsymbol{\beta}}_{h,\mathsf{Q}}-\boldsymbol{\beta}^*\|_2/C\leq&\gamma^{1/2}h^2\vee\sqrt{\frac{\xi_t}{n|\mathcal{E}|}\gamma}\vee\gamma\sqrt{|S^*|}\frac{t+\log(|\mathcal{E}||{S^*}|)}{n\min_{j\in S^*}|\beta^*_j|}\\
        &\vee \gamma\frac{\xi_t^{1/2}(|S^*|+t+\log(2|\mathcal{E}|))^{1/2}}{nh^{1/2}}\vee\gamma\frac{\xi_t(|S^*|+t+\log(2|\mathcal{E}|))^{1/2}}{n^{3/2}h}\\
        &\vee\gamma\left(\frac{\xi_th^3}{n}\right)^{1/2}\vee\gamma\frac{\xi_th}{n}\vee\gamma^{1/2}h\left(\frac{|{S^*}|+t+\log|\mathcal{E}|+1}{n}\right)^{1/4}:=\kappa
    \end{align*}
    for some universal constant $C$ that only depends on $(l_0,k_2,\sigma_x,\lambda_{U},\lambda_L,k_u,k_l,f_u,f_l).$
\end{theorem}\vspace{-0.07cm}
\begin{theorem}[Non-asymptotic Variable Selection Property]\label{thm:non-asymptotic Variable Selection Consistency}
    Under the conditions of Theorem \ref{thm:non-asymptotic ell_2 Error Bound}, define $s_+=\min_{j\in S^*}|\boldsymbol{\beta}_j^*|^2$, we have $S^*\subseteq\mathrm{supp}(\hat{\boldsymbol{\beta}}_{h,\mathsf{Q}})\subseteq (G_{\boldsymbol{\omega}})^c$ with probability at least $1-10e^{-t}$ if $\kappa^2\lesssim s_+$, where $\kappa$ is defined in Theorem \ref{thm:non-asymptotic ell_2 Error Bound}.
\end{theorem}
Theorem \ref{thm:non-asymptotic ell_2 Error Bound} allows some flexibility in the choice of bandwidth $h$. For example, assume further that $C_{n,p}=O(1)$, $s_-\gtrsim1$, and $\gamma\asymp1$ with $\gamma\geq\gamma^*$, $(\xi_t/(n|\mathcal{E}|))^{1/2}\vee(|\mathcal{E}|\xi_t(|S^*|+t+\log(2|\mathcal{E}|)))^{1/2}/n\vee(|S^*|+t+\log(2|\mathcal{E}|))|\mathcal{E}|/n\lesssim h\lesssim(\xi_t/(n|\mathcal{E}|^2))^{1/4}\wedge(1/|\mathcal{E}|)^{1/3}$, and the $\beta$-min condition $\min_{j\in S^*}|\beta^*_j|\gtrsim|S^*|^{1/2}(t+\log(|\mathcal{E}||{S^*}|))\cdot(|\mathcal{E}|/(n\xi_t))^{1/2}$, we conclude that $\|\hat{\boldsymbol{\beta}}_{h,\mathsf{Q}}-\boldsymbol{\beta}^*\|_2\lesssim(\xi_t/(n|\mathcal{E}|))^{1/2}$, which aligns with the convergence rate in \citet{fan2024environment} up to logarithmic factors. This provides a wide range for the choice of $h$. Under the choice of $h$ above, we have the variable selection property if $\xi_t/(n|\mathcal{E}|)\lesssim s_+\wedge (\gamma s_-)$. In practice, we use $h=\sqrt{\tau(1-\tau)p\gamma/n_*}$ as a numerical default, where $n_*=\min_{e\in\mathcal{E}}n^{(e)}/\omega^{(e)}$. Section B.2 examines bandwidth sensitivity of our method.

\vspace{-0.3cm}

\section{Illustration and Numerical Experiments}\label{sec5}\vspace{-0.2cm}
\subsection{Simulations}\label{sec5.1}\vspace{-0.16cm}
Throughout this section, we fit KS-IQR with an intercept and without imposing an explicit bound on $\|\boldsymbol{\beta}\|_2$ (see details in Section B.1 in the Supplementary Material). Our first comparison is conducted against the EILLS estimator \citep{fan2024environment}. We begin with the following structural causal model (SCM).

\textbf{Model 1.} We consider two environments with equal sample sizes $n$. The structural assignments of the SCM in $e=1$ and $e=2$ are exhibited below
\begin{align*}
    x^{(e)}_1 &\gets u_1^{(e)},\qquad  \qquad \qquad \qquad x^{(1)}_4 \gets u_4^{(1)},\qquad  \qquad \qquad \qquad x^{(2)}_4 \gets (u_4^{(2)})^2-1,\\
    x^{(e)}_2 &\gets \sin(x_4^{(e)}) + u_2^{(e)}, \qquad x^{(e)}_3 \gets \cos(x_4^{(e)}) + u_3^{(e)},\qquad x^{(e)}_5 \gets \sin(x_3^{(e)} + u_5^{(e)}), \\
    x^{(e)}_{10} &\gets 2.5 x_1^{(e)} + 1.5 x_2^{(e)} + u_{10}^{(e)},\qquad \qquad \qquad y^{(e)} \gets 3 x_1^{(e)} + 2x_2^{(e)} - 0.5 x_3^{(e)} + u_{13}^{(e)}, \\
    x_6^{(e)} &\gets 0.8 y^{(e)} u_6^{(e)}, \\
    x_7^{(1)} &\gets 0.5 x_3^{(1)} + y^{(1)} + u_7^{(1)}, \qquad \qquad \qquad \qquad x_7^{(2)} \gets 4 x_3^{(2)} + \tanh(y^{(2)}) + u_7^{(2)}, \\
    x_8^{(e)} &\gets 0.5 x_7^{(e)} - y^{(e)} + x_{10}^{(e)} + u_8^{(e)}, \qquad \qquad x_9^{(e)} \gets \tanh(x_7^{(e)}) + 0.1 \cos(x_8^{(e)}) + u_9^{(e)}, \\
    x^{(e)}_{11} &\gets 0.4 (x_7^{(e)} + x_8^{(e)}) * u_{11}^{(e)}, \qquad \qquad \qquad \qquad x^{(e)}_{12} \gets u_{12}^{(e)},
\end{align*} 
where $u^{(e)}_1, \ldots, u^{(e)}_{13}$ are independent random variables. Consider two cases: \textbf{(i)} $(u^{(e)}_1, \ldots, u^{(e)}_{13})^\top \sim \mathcal{N}(\boldsymbol{0}, \boldsymbol{I}_{13\times 13})$ for $e\in\{1,2\}$; \textbf{(ii)} $u^{(1)}_1,\ldots,u^{(1)}_{13} \sim t_{1.5}$, and $(u^{(2)}_1,\ldots,u^{(2)}_{13})^\top\sim \mathcal{N}(\boldsymbol{0}, \boldsymbol{I}_{13\times 13})$.

Model~1 under case (i) is the same as the model in Section 5 of \citet{fan2024environment}, whereas Model~1 under case (ii) introduces heavy-tailed variables in the first environment. The true parameter is $\boldsymbol{\beta}^*=(3,2,-0.5,0,\ldots,0)^\top$ and $S^*=\{1,2,3\}$. The endogenous variables are indexed by $G=\{7,8,9\}$ under the conditional mean setting of \citet{fan2024environment}, and the same endogenous structure is preserved under IQR framework at the median in both cases.

\begin{figure}[!t]
    \begin{subfigure}{\textwidth}
        \centering
        \begin{subfigure}{0.265\textwidth}
            \centering
            \includegraphics[width=\textwidth]{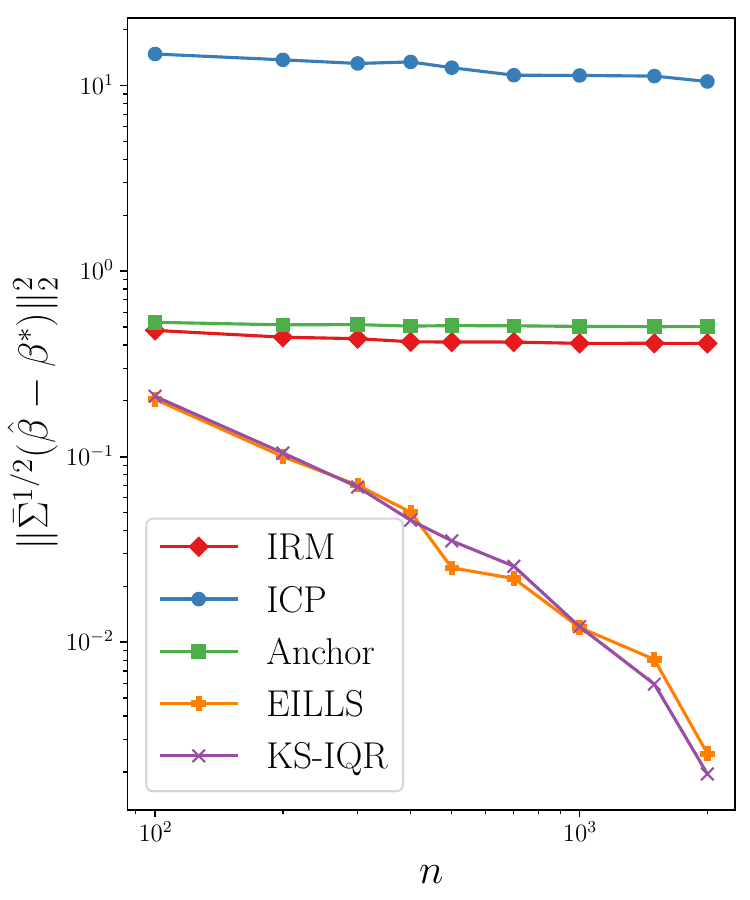}
            \caption{(a)}
            \label{fig1a}
        \end{subfigure}
        \hfill
        \begin{subfigure}{0.265\textwidth}
            \centering
            \includegraphics[width=\textwidth]{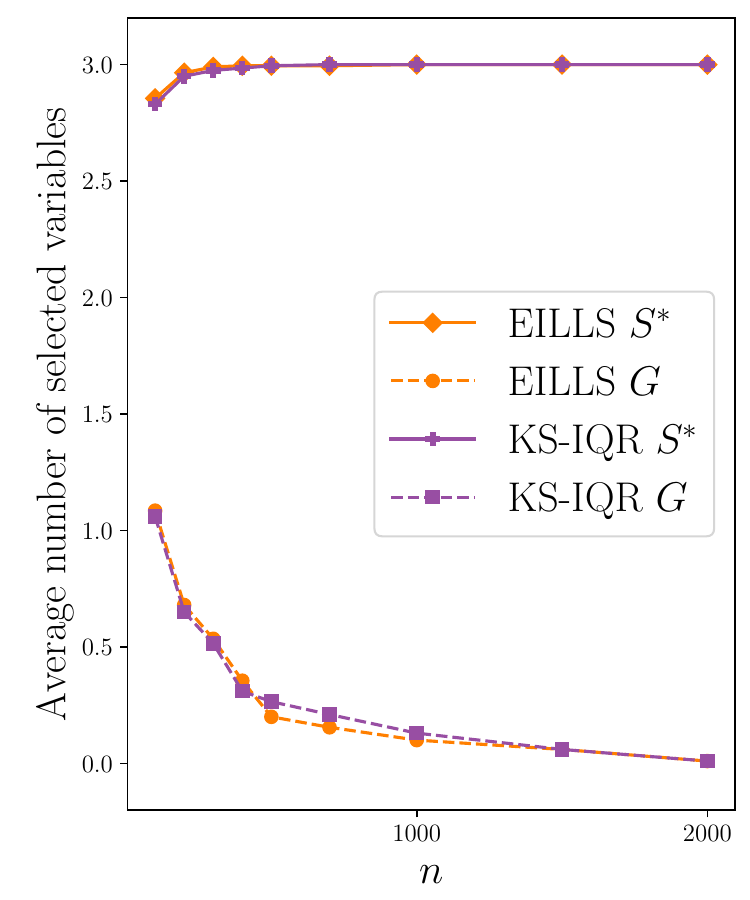}
            \caption{(b)}
            \label{fig1b}
        \end{subfigure}
        \hfill
        \begin{subfigure}{0.265\textwidth}
            \centering
            \includegraphics[width=\textwidth]{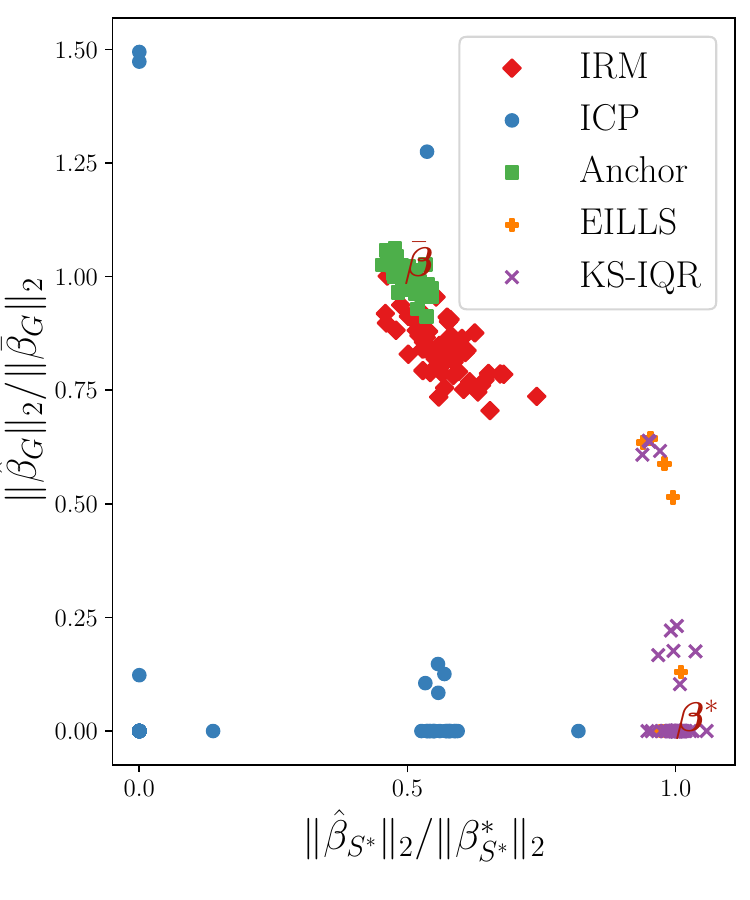}
            \caption{(c)}
            \label{fig1c}
        \end{subfigure}
    \end{subfigure}

    \par
    \centering
    Case (i)
    \par
    \vspace{-0.1cm}

    \begin{subfigure}{\textwidth}
        \centering
        \begin{subfigure}{0.265\textwidth}
            \centering
            \includegraphics[width=\textwidth]{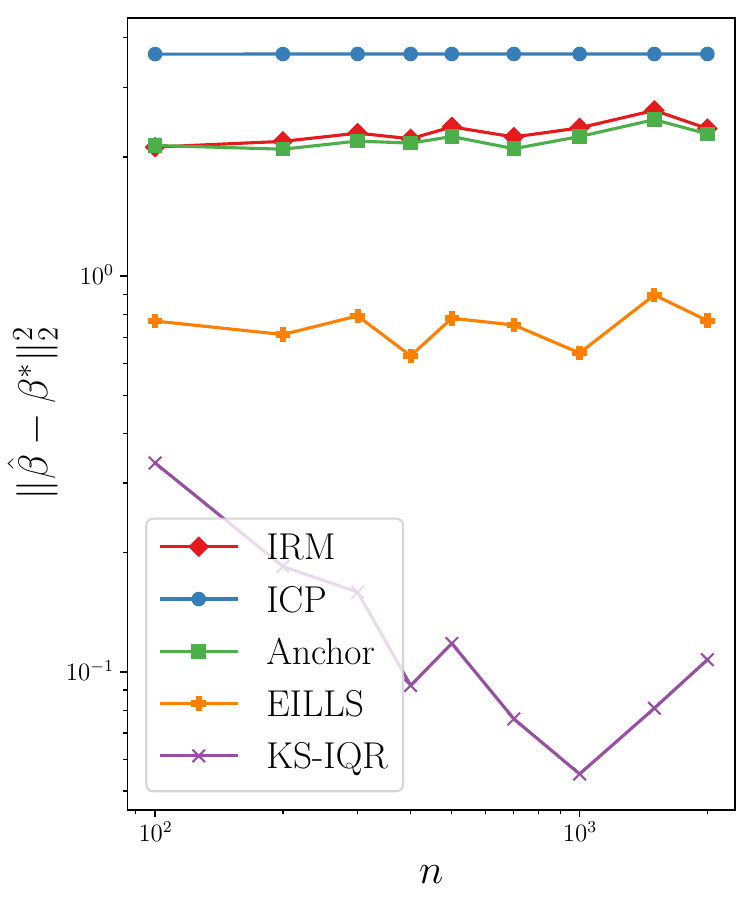}
            \caption{(d)}
            \label{fig1d}
        \end{subfigure}
        \hfill
        \begin{subfigure}{0.265\textwidth}
            \centering
            \includegraphics[width=\textwidth]{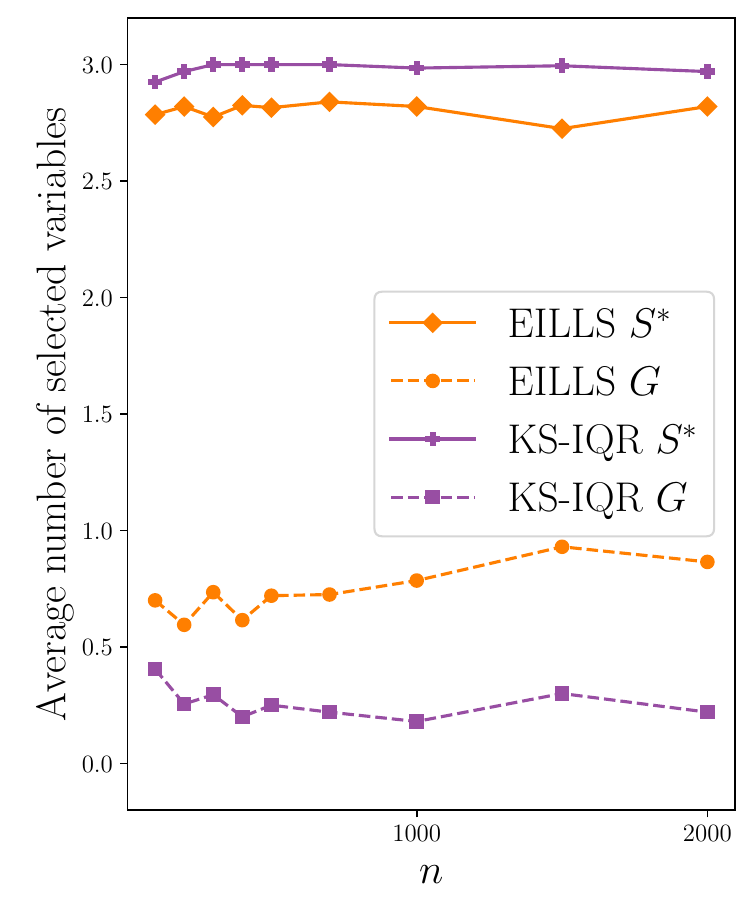}
            \caption{(e)}
            \label{fig1e}
        \end{subfigure}
        \hfill
        \begin{subfigure}{0.265\textwidth}
            \centering
            \includegraphics[width=\textwidth]{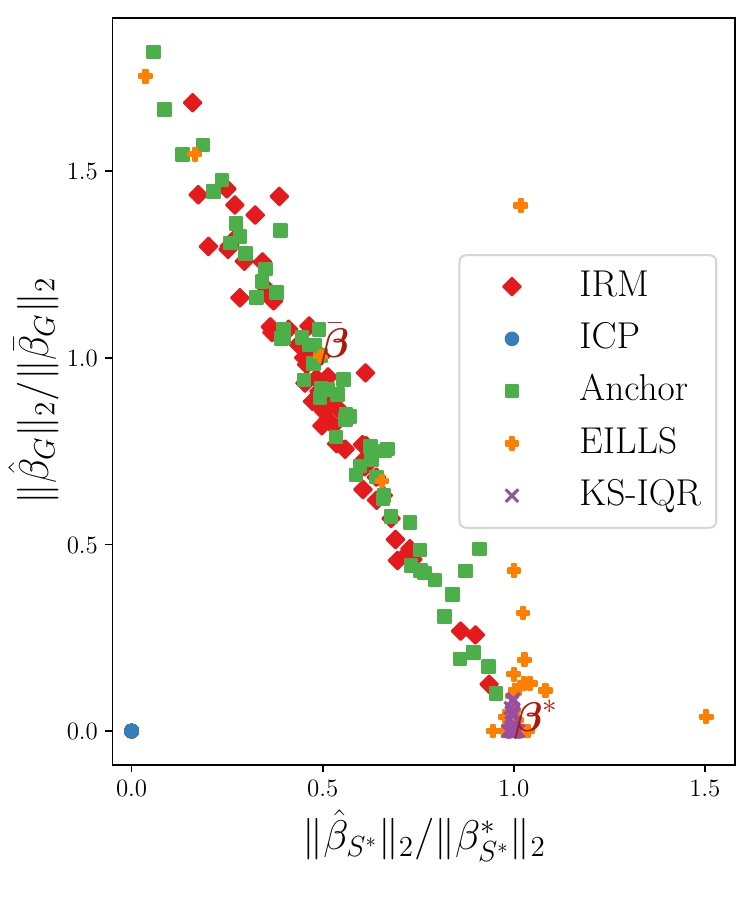}
            \caption{(f)}
            \label{fig1f}
        \end{subfigure}
    \end{subfigure}
    \par
    \centering
    Case (ii)
    \par
    \vspace{-0.1cm}
    \captionsetup{font=footnotesize}
    \caption{Simulation results for Model 1, where the first and second rows correspond to Model 1 under case (i) and Model 1 under case (ii), respectively. (a) reports the averaged $\ell_2$ error $\|\bar{\boldsymbol{\Sigma}}^{1/2}(\hat{\boldsymbol{\beta}}-\boldsymbol{\beta}^*)\|_2^2$ while (d) reports the averaged $\ell_2$ error $\|\hat{\boldsymbol{\beta}}-\boldsymbol{\beta}^*\|_2^2$ (since $\bar{\boldsymbol{\Sigma}}$ is unstable in Model 1 under case (ii)) over $200$ replications for each approach as a function of sample size $n$. (b) and (e) show the average number of selected variables from the true support $S^*=\{1,2,3\}$ and the endogenous set $G=\{7,8,9\}$ across $200$ replications under varying $n$ for the EILLS and KS-IQR estimators. (c) and (f) provide a visual comparison of the solutions produced by each method over $60$ repeated trials at $n=700$. We show in red the true parameter $\boldsymbol{\beta}^*$ and the pooled least-squares estimate $\bar{\boldsymbol{\beta}}$ computed from $10^6$ observations with Gaussian noises per environment.}
    \label{fig1}
\end{figure}

For the EILLS estimator and our KS-IQR estimator, we set the hyper-parameter $\gamma=20$ and balanced weights $\omega^{(1)}=\omega^{(2)}=1/2$, and compute both estimators via brute force. For the KS-IQR estimator, we fix the quantile level at $\tau=0.5$, employ the Gaussian kernel, and choose the bandwidth as $h=\sqrt{\tau(1-\tau)p\gamma/n_*}$, where $n_*=\min_{e\in\mathcal{E}}n^{(e)}/\omega^{(e)}=2n$. We do not claim such a choice of $h$ is numerically optimal. Further discussion regarding the choice of $h$ is provided in Section B.2 of the Supplementary Material. We also employ IRM \citep{arjovsky2019invariant}, anchor regression (Anchor) \citep{rothenhausler2021anchor}, invariance causal prediction (ICP) \citep{peters2016causal} as baselines. The hyper-parameters governing the invariance penalties in IRM, Anchor, and ICP are determined via oracle selection: each candidate value is evaluated exhaustively, and the one yielding the minimum $\ell_2$ error $\|\bar{\boldsymbol{\Sigma}}^{1/2}(\hat{\boldsymbol{\beta}}-\boldsymbol{\beta}^*)\|_2^2$ is adopted, where $\bar{\boldsymbol{\Sigma}}$ is the empirical sample covariance matrix. We set the sample size $n\in\{100,200,300,400,500,700,1000,1500,2000\}$. 
The results are exhibited in Figure \ref{fig1}. For Model~1 under case (i), as shown in Figure~\ref{fig1a}, EILLS and KS-IQR are the only two methods that consistently estimate $\boldsymbol{\beta}^*$. This observation is further confirmed by Figure~\ref{fig1c}. The number of selected variables from $S^*$ and $G$ presented in Figure~\ref{fig1b} is also highly similar between these two methods, illustrating that their performances are comparable in this setting. However, when it comes to Model~1 under case (ii), the EILLS estimator fails to converge as $n$ increases. This is evident from the scatter plot in Figure~\ref{fig1f}, which indicates that EILLS becomes unstable in this case. In contrast, the KS-IQR estimator remains close to the true parameter $\boldsymbol{\beta}^*$. From Figure~\ref{fig1e}, it can also be observed that EILLS selects fewer variables from $S^*$ and more variables from $G$, implying that EILLS fails to correctly conduct causal discovery and exclude endogenous variables. 
These results also demonstrate the empirical effectiveness of unconstrained KS-IQR.

\begin{figure}[!t]
\centering
\hspace{0.11\textwidth}
    \begin{subfigure}{0.32\textwidth}
        \centering
        \includegraphics[width=\textwidth]{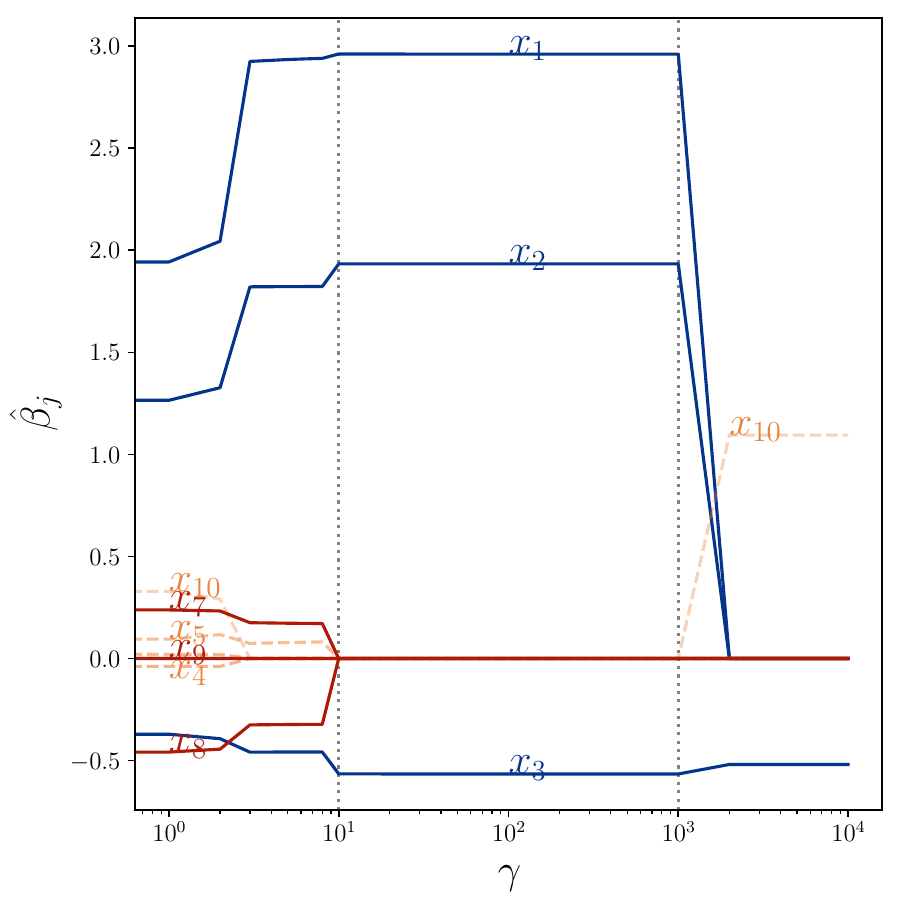}
        \caption{(a)}
        \label{fig2a}
    \end{subfigure}
    \hfill
    \begin{subfigure}{0.27\textwidth}
        \centering
        \includegraphics[width=\textwidth]{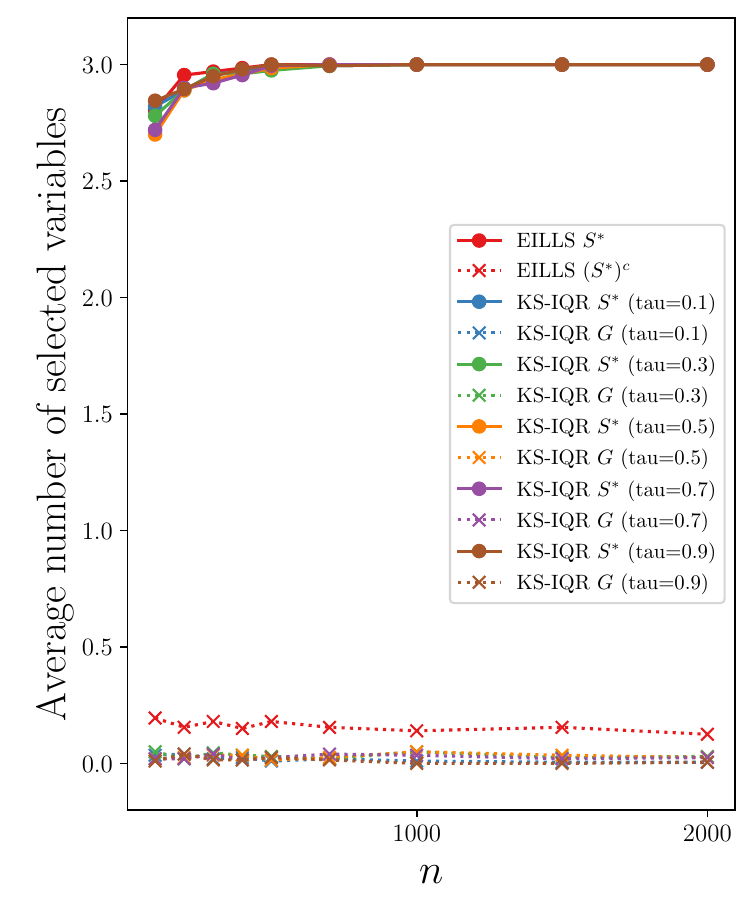}
        \caption{(b)}
        \label{fig2b}
    \end{subfigure}
    \hspace{0.11\textwidth}
    \captionsetup{font=footnotesize}
    \caption{(a): Solution paths of the KS-IQR estimator (with $\tau=0.5,~h=0.1$, and a Gaussian kernel) in a single trial under Model~1~(i) as $\gamma$ varies. The vertical axis corresponds to $\hat{\beta}_j$, the $j$-th component of $\hat{\boldsymbol{\beta}}_{h,\mathsf{Q}}$. (b): Average number of selected variables indexed by the true support $S^*=\{1,2,3\}$ and the non-parent set $G=\{4\}$ across $200$ replications under varying $n$ for the KS-IQR (quantile levels $\tau\in\{0.1, 0.3, 0.5, 0.7, 0.9\}$, bandwidth $h=\sqrt{\tau(1-\tau)p\gamma/n_*}$, Gaussian kernel) and the EILLS estimators under Model 2.}
    \label{fig2}
\end{figure}

We further conduct experiments to examine how the KS-IQR estimator $\hat{\boldsymbol{\beta}}_{h,\mathsf{Q}}$ is influenced by the penalty parameter $\gamma$. We consider Model~1 under case (i) and fix $n=300$, $\tau=0.5$, $h=0.1$, and report each component of $\hat{\boldsymbol{\beta}}_{h,\mathsf{Q}}$ under varying $\gamma$. The results are presented in Figure~\ref{fig2a}. It can be observed from Figure~\ref{fig2a} that when $\gamma\in[10,1000]$, KS-IQR precisely recovers the support set $\{1,2,3\}$. In this regime, it includes all variables indexed by $S^*$ and screens out all endogenous variables indexed by $G$, while completely eliminating the exogenous variables indexed by $[12]\setminus(S^*\cup G)$. Setting $\gamma=0$ reduces KS-IQR to pooled smoothed quantile regression, which selects all variables, further highlighting the importance of incorporating the penalty term. When $\gamma>1000$, KS-IQR tends to select $\{3,10\}$ rather than $\{1,2,3\}$, illustrating the necessity of including the smoothed quantile loss in the objective function, as it becomes negligible when $\gamma$ is sufficiently large. Nevertheless, the results suggest that the estimator is stable over a relatively wide range of $\gamma$, indicating that the method is not overly sensitive to the choice of $\gamma$.

To assess the set of endogenous variables, we consider the following model.

\textbf{Model 2.} We consider two environments with equal sample sizes $n$. The structural assignments of the SCM in $e=1$ and $e=2$ are given below
\begin{align*}
    x_1^{(e)} \gets u_{1}^{(e)},~~ x_3^{(e)} \gets u_{3}^{(e)},~~ x_2^{(e)} &\gets \sin(x_3^{(e)}) + u_{2}^{(e)},~~ y^{(e)} \gets 3 x_1^{(e)} + 2x_2^{(e)} - 0.5 x_3^{(e)} + u_{5}^{(e)}, \\
    x_4^{(1)} &\gets 5(y^{(1)})^2 + u_{4}^{(1)}, \quad x_4^{(2)} \gets u_4^{(2)},
\end{align*} 
where $(u^{(e)}_1,\ldots,u^{(e)}_{5})^\top\sim \mathcal{N}(\boldsymbol{0}, \boldsymbol{I}_{5\times 5})$ for all $e\in\{1,2\}$.

For Model~2, the true quantile intercept is $\Phi^{-1}(\tau)$ and the coefficient vector is $\boldsymbol{\beta}^*=(3,2,-0.5,0)^\top$, with true support $S^*=\{1,2,3\}$. The non-parent $x_4$ is not classified as endogenous under the conditional mean setting of \citet{fan2024environment}, since $\mathbb{E}[x_4^{(1)}u_5^{(1)}]=\mathbb{E}[x_4^{(2)}u_5^{(2)}]=0$. Under equal environment weights, its pooled quantile score is nonzero for $\tau\neq 0.5$. Recall that one of our goals is to conduct causal discovery, but the circumvention of spurious variables is also important beyond incorporating all important variables. This highlights the merit of applying IQR rather than focusing only on the conditional mean. Now we examine the number of selected variables from the true support $S^*=\{1,2,3\}$ and the non-parent set $G=\{4\}$. To provide a comprehensive comparison between the EILLS and KS-IQR estimators, we consider quantile levels $\tau\in\{0.1,0.3,0.5,0.7,0.9\}$, fix the hyper-parameter $\gamma=20$, and set the bandwidth as $h=\sqrt{\tau(1-\tau)p\gamma/n_*}$ using Gaussian kernel as before. We then report the number of variables selected by each method under varying sample sizes $n\in\{100,200,300,400,500,700,1000,1500,2000\}$. Figure~\ref{fig2b} shows that KS-IQR selects $x_4$ less frequently than EILLS at the quantile levels considered. At larger sample sizes, both methods correctly select variables from $S^*$.

\textbf{Model 3.} We consider two equally weighted training environments. In each, let $Z$ and $W$ be independent Rademacher variables and let $U\sim\mathbb{N}(0,1)$ be independent of $(Z,W)$. In environment $e$, we generate $Y=1+2Z+(1+W/2)U,~\mathbb{P}_e(D=d|Z,W,U)=\{1+d\,a_e\tanh(U)\}/2,~ d\in\{-1,1\}$. The conditional $\tau$-quantile of $Y$ given $(Z,W)$ is $1+2Z+(1+W/2)\Phi^{-1}(\tau)$, where $\Phi$ is the standard normal distribution function. Thus $W$ enters the upper quantile but not the median, whereas $D$ is an endogenously spurious variable. We take $a_1$ from $0.50$ to $0.80$ with step size $0.05$, $a_2=1-a_1$, and define $a_c=|a_1-a_2|$ as the contrast. For the Gaussian kernel, fixed $h>0$, and a fixed parameter ball containing $\boldsymbol{\beta}^*$, Assumption \ref{assum4.5} holds whenever $a_c>0$, since the two environments' population scores for $D$ differ uniformly on the ball. Since $a_1+a_2=1$ throughout, these choices keep the pooled distribution of a single observation the same across all three cases, while increasing the contrast between environments. We evaluate prediction in two new environments, $a=0$ and $a=-0.8$, where the association between $D$ and $U$ disappears and reverses, respectively. We retain the tuning settings of Model~2 for KS-IQR and EILLS; pooled SQR uses the same kernel and bandwidth as KS-IQR, with $\gamma=0$ in the penalty.
\begin{figure}[!t]
    \centering
    \includegraphics[width=0.96\textwidth]{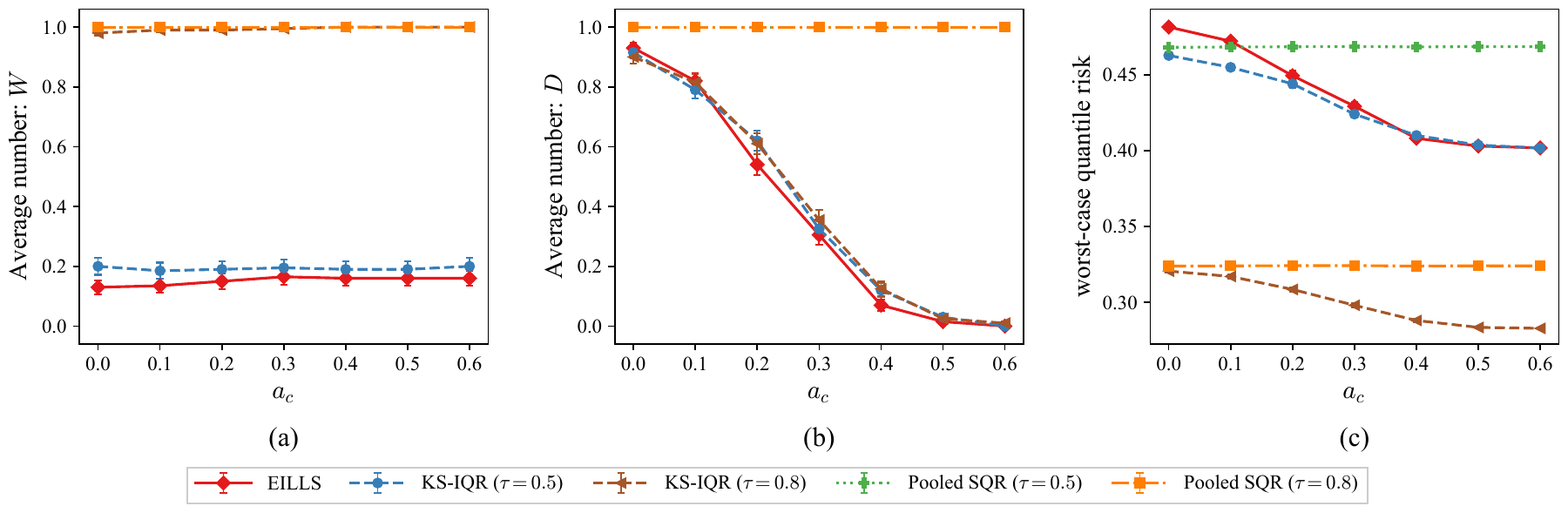}
    \captionsetup{font=footnotesize}
    \caption{Variable selection and worst-case prediction under quantile risk across training environment contrasts in Model 3 ($n=500$ per training environment; $200$ replications). (a) and (b) show the average number of selected $W$ and $D$. (c) shows the larger of the two test-environment check losses in each replication, averaged over replications. Each test environment contains $10,000$ observations. EILLS predictions are evaluated at the median. Error bars show one Monte Carlo standard error.}
    \label{fig4}
\end{figure}
Figure~\ref{fig4} compares KS-IQR and pooled SQR at $\tau=0.5$ and $0.8$ for $n=500$ per training environment, with EILLS as a conditional mean comparison. At the median, KS-IQR often missed $W$, a structural parent that affects the distribution of $Y$ but not its conditional median; EILLS also selects $W$ infrequently. KS-IQR has a lower worst-case quantile risk on the two test environments than EILLS under no or weak training contrast. When $a_c=0.6$, both methods exclude $D$ and have nearly identical losses. At $\tau=0.8$, KS-IQR and pooled SQR nearly always select $W$, illustrating why examining an upper quantile can reveal a parent that a median analysis often misses. As the contrast increases, the selection frequency of $D$ decreases for KS-IQR, whereas pooled SQR cannot exclude $D$. Moreover, KS-IQR exhibits lower worst-case quantile risk than pooled SQR. These results are also consistent with the robust prediction interpretation of Proposition \ref{prop: robust oos quantile risk}: the two test environments preserve the conditional quantile given $(Z,W)$ while changing the association between $D$ and the noise. Section B.3 reports results for other sample sizes. Section B.4 presents a separate simulation in which a variable's effect depends on the latent noise.\vspace{-0.35cm}
\subsection{Real Data Analysis}\label{sec5.2}\vspace{-0.15cm}
In this section, we apply our method to the Flow Cytometry Data \citep{sachs2005causal}, a standard benchmark for causal discovery in biological systems\footnote{The code for replicating results in Section 5 is available at \href{https://github.com/BoFuxjtu/IQR.git}{https://github.com/BoFuxjtu/IQR.git}.}. The dataset is publicly available as supplementary material to \cite{sachs2005causal} at \url{https://www.science.org/doi/suppl/10.1126/science.1105809/suppl_file/sachs.som.datasets.zip}. The dataset records the abundance of $11$ proteins and phospholipids under multiple experimental conditions, including one baseline (unperturbed) condition and several interventions induced by reagents (or stimulant removals). Each environment contains approximately $700$--$1000$ samples. Following \citet{mooij2013cyclic,meinshausen2016methods}, we restrict attention to a subset of eight environments. These interventions are designed to perturb specific signaling pathways, although their effects may not be perfectly targeted, and off-target influences can occur. Consequently, interventions may also affect the response variable, leading to uncertainty in the true intervention mechanism. To mitigate this issue, we follow \citet{meinshausen2016methods,peters2016causal} and perform pairwise analysis across all 28 environment pairs. We compare our KS-IQR with EILLS \citep{fan2024environment}, ICP \citep{peters2016causal}, hiddenICP \citep{meinshausen2016methods}, Cyclic Causal Discovery (CCD; \citealp{mooij2013cyclic}), Bayesian structure learning (BSL; \citealp{pmlr-v2-eaton07a}). We also include the consensus network (Sachs-a; \citealp{sachs2005causal}) and the author's reconstructed network (Sachs-b; \citealp{sachs2005causal}) as benchmarks. We standardize each variable and the response separately within each environment. For the EILLS estimator and our KS-IQR estimator, we fix the hyper-parameter $\gamma=20$ and use balanced weights $\omega^{(1)}=\omega^{(2)}=1/2$, with estimators obtained via brute force. For KS-IQR, we adopt a Gaussian kernel with bandwidth $h=0.1$ and evaluate quantile levels $\tau\in\{0.1,0.2,0.3,0.4,0.5,0.6,0.7,0.8,0.9\}$.

In \citet{meinshausen2016methods}, the authors consider the union formed by the inferred parental sets across all 28 environment pairs. Therefore, we aggregate results across environment pairs via support frequencies for EILLS and KS-IQR. Specifically, estimated coefficients with magnitude exceeding $10^{-2}$ are regarded as indicative of a causal relationship. For EILLS, a relationship is retained if it is supported by at least $50\%$ of the environment pairs. For KS-IQR, the support ratio is computed at each quantile level; a causal relationship is then retained if it satisfies either criterion (i) or criterion (ii): (i) it exceeds $60\%$ at a specific quantile level, or (ii) its average across all given quantile levels exceeds $30\%$. We adopt these criteria to accommodate heterogeneity across quantile levels: some relationships may manifest only at specific (e.g., tail) quantiles, whereas others remain consistent across quantiles. We summarize the identified causal relationships in Table B.1 and provide complete results for all 110 candidate relationships in Table B.2 of the Supplementary Material, including selection frequencies across all environment pairs and quantile levels.

From Table B.1, we observe a substantial degree of consistency among the results obtained by different methods. For example, the two-cycles AKT$\leftrightarrows$ERK and RAF$\leftrightarrows$MEK are identified by both EILLS and KS-IQR. The two-cycle PIP3$\leftrightarrows$PLCg is also identified by KS-IQR. Notably, EILLS is the only method that fails to identify the relationship PKC $\to$ JNK. In addition, several causal relationships uniquely discovered by KS-IQR are supported by existing biological evidence. For example, the identified ERK$\to$MEK relationship is consistent with the negative feedback phosphorylation of MEK in the ERK1/2 MAPK pathway described by \citet{lake2016negative}. Similar findings are reported in \citet{dougherty2005regulation,shin2009positive}. Furthermore, ERK is known to interact with the PI3K/AKT pathway (although PI3K is not observed in this dataset), thereby influencing PIP2 and PIP3 \citep{zmajkovicova2013mek1}, which supports the edges ERK$\to$PIP2 and ERK$\to$PIP3. Moreover, P38 has been shown to be dependent on the activation of ERK \citep{poddar2013novel}, supporting the edge ERK$\to$P38. The activation of JNK is modulated by the ERK and AKT pathways \citep{fey2012crosstalk}, supporting the edges ERK$\to$JNK and AKT$\to$JNK. These findings underscore the advantage of KS-IQR in capturing a broader set of candidate causal relationships. In particular, KS-IQR retained AKT$\to$JNK under Criterion (ii), with mean support of $42.9\%$ across the $9$ quantiles, whereas none of the other methods retained this edge. Additionally, as shown in Table B.2, using a threshold of at least $50\%$ pairs, median KS-IQR retained nine directed edges, all among the 12 retained by EILLS: MEK$\to$RAF, PIP3$\to$PIP2, PIP2$\to$PIP3, AKT$\to$ERK, ERK$\to$AKT, AKT$\to$PKA, P38$\to$PKC, PKC$\to$P38, and JNK$\to$P38. Across $9$ quantiles, KS-IQR retained $33$ edges under the primary rule and $28$ when the cutoff in criterion~(ii) is raised from $30\%$ to $40\%$. Of the $12$ edges with the highest average support, $11$ are also retained by EILLS.\vspace{-0.55cm}
\section{Conclusion}\label{sec6}\vspace{-0.2cm}
In this paper, we consider an invariant quantile regression model in a multi-environment setting. The proposed framework enables the identification of invariant causal variables that remain stable across environments while capturing heterogeneous causal relationships along the outcome distribution. To implement this idea, we develop a KS-IQR estimator and establish its theoretical properties for global minimizers over a bounded parameter space. Under Assumption \ref{assum4.5}, the estimator consistently recovers invariant causal variables while excluding endogenous variables, together with a non-asymptotic $\ell_2$ error bound. Our analysis further shows that quantile-based invariance provides a richer characterization of the conditional distribution of the outcome, with direct implications for causal discovery and particular relevance to settings where tail behavior is of primary interest. Moreover, compared with conditional-mean approaches, the proposed framework more effectively resolves endogenous structures and therefore enables stricter elimination of spurious (non-causal) variables. Several directions merit further investigation, including extending the proposed framework to high-dimensional settings by considering the approaches in \citet{gu2025fundamental}, which assume additional structural conditions, and developing statistical inference procedures for quantile-based invariant causal relationships.

\vspace{-0.6cm}

\section*{Disclosure Statement}\label{disclosure-statement}\vspace{-0.1cm}
The authors do not have any relevant financial or non-financial competing interests.\vspace{-0.1cm}

\noindent\textit{Generative AI Use.}\vspace{-0.1cm}
The authors used OpenAI ChatGPT (GPT-6) to assist with language refinement and reviewing mathematical derivations and proofs. The authors independently verified the content, checked the tool's terms of use, and take full responsibility for the originality, accuracy, and integrity of the manuscript.

\vspace{-0.57cm}

\section*{Acknowledgement}\label{Acknowledgement}\vspace{-0.15cm}
The authors were supported by NSFC under Grant No. 12571311.

\vspace{-0.5cm}

\section*{Supplementary Material}\vspace{-0.15cm}
The supplementary material contains proofs, auxiliary lemmas, and additional numerical experiments.

\vspace{-0.65cm}
\bibliography{bibliography.bib}

\end{document}